\documentstyle[dina4,12pt]{myart}

\newcommand{\be}{\begin{equation}}
\newcommand{\ee}{\end{equation}}
\newcommand{\bea}{\begin{eqnarray}}
\newcommand{\eea}{\end{eqnarray}}
\newcommand{\nn}{\nonumber\\}

\newcommand{\ubr}{\bar{u}}
\begin{document}
\def \L {\cal L}
\title{Constraint Correlation Dynamics of  SU(N) Gauge Theories\footnote{Part
of the dissertations of A. Peter and J.M. H\"auser}
\thanks{This work was supported in part
by the National Natural Science Foundation and the Doctoral Education
Fund of the State Education Commission of China,
by the Deutsche Forschungsgemeinschaft and GSI Darmstadt.}\bigskip}
\author{S.J. Wang$^{1,2}$, W. Cassing$^1$, J.M. H\"auser$^1$, A. Peter$^1$
 and M.H. Thoma$^1$ \\
$^1${\it Institut f\"ur Theoretische Physik, Universit\"at Giessen,}
\\{\it 35392 Giessen, Germany }  \\
$^2${\it Department of Modern Physics, Lanzhou University,}\\
{\it Lanzhou 730000, PR China }}

\maketitle

\begin{abstract}
A constraint correlation dynamics up to 4-point Green functions is proposed
for SU(N) gauge theories which reduces the N-body quantum field problem to
the two-body level. The resulting set of nonlinear coupled equations fulfills
all conservation laws including fermion number, linear and angular momenta
as well as the total energy. Apart from the conservation laws in the space-time
degrees of freedom the Gauss law is conserved as a quantum expectation
value identically for all times. The same holds for the Ward identities as
generated by commutators of Gauss operators. The constraint dynamical
equations are highly non-perturbative and thus applicable also in the strong
coupling regime, as e.g. low-energy QCD problems.
\end{abstract}
\newpage

\zeqn
\section{Introduction}

A lot of problems in QCD at zero temperature, finite temperature, and in
the case of non-equilibrium situations require the use of non-perturbative
methods. Important examples are the mechanism of confinement, probably
related to the vacuum structure \cite{SHU}, the investigation of hadron
spectra and hadronic matter, the understanding of the nature of the
deconfinement transition and the chiral symmetry restoration.
Furthermore, the mechanism of hadronization in relativistic heavy-ion
collisions, -- following the possible preformation of a quark-gluon plasma --
or possible non-perturbative effects in a quark-gluon plasma above the critical
temperature cannot be treated perturbatively as well \cite{r1a}.
Besides numerically
very involved lattice calculations \cite{r1} there are only preliminary
attempts to incorporate non-perturbative effects by using the Dyson-Schwinger
equation or variational calculations \cite{r2} - \cite{r2c} \ .
In addition, considering the formation, evolution,
and decay of a quark-gluon plasma
in relativistic heavy-ion collisions, a transport theory based on gauge
covariant Wigner functions has been proposed by Elze et al.
\cite{r3a,r3}. It is, in
fact, a many-body theory completely equivalent to the Heisenberg equations
for the quarks and gluon fields, however, as a reformulation of the SU(N)
gauge theory in phase space not manageable in practice. The standard
applicable limit is the semi-classical, abelian approximation which
corresponds to lowest order perturbation theory in the high temperature
approximation. A more promising approach may be provided by the use of
equal-time Wigner functions \cite{r4}.

In this paper we propose a different approach along the line
of relativistic two-body correlation dynamics \cite{r5}-\cite{r9}, which
has proven to provide the genuine basis for the formulation of non-perturbative
transport theories for baryons and mesons \cite{r8}. It is already
known that this approach obeys all conservation laws concerning space-time
degrees of freedom, i.e. fermion number, linear and angular momentum as well
as the total energy \cite{r5}. The novel phenomena in the application to
quarks and gluons is the non-abelian dynamics of the gluon fields as well as
a SU(N) internal gauge symmetry.

The paper is organized as follows: In Section 2 we first give a brief reminder
of the SU(N) gauge theory in the temporal gauge and introduce the necessary
notations.
Section 3 is devoted to the derivation of equations
of motion and suitable truncation schemes for linked $n$-point Green functions.
In Section 4 we show the conservation of the Gauss law in time (as a
quantum expectation value) within the
approximation scheme adopted and investigate the higher order Ward identities
(as generated by commutators of Gauss operators),
whereas Section 5 provides a summary
and discussion of open problems. The explicit form of the final constraint
dynamical equations is in part shifted to the Appendices in view
of their length.

\newpage

\zeqn
\section{SU(N) gauge theories in the temporal gauge}

Here we repeat the basic concepts and equations of SU(N) gauge theories
\cite{r10}, which build the starting point for the correlation dynamics in QCD.

We begin with the SU(N) gauge Lagrangian
\begin{equation}
{\cal L} = - \bar{\Psi} \gamma_{\mu} D_{\mu} \Psi
 - \frac{1}{4} F^a_{\mu \nu} F^a_{\mu \nu},
\label{2.1}
\end{equation}
where the gauge-covariant derivative $D_{\mu}$ is defined as
\begin{equation}
D_{\mu} = \partial_{\mu} - i g T^a A^a_{\mu} \ .
\label{2.2}
\end{equation}
The fields $A^a_{\mu}$ are the gauge potentials and $T^a$ are the
$N \times N$ generators of SU(N) following
\begin{equation}
T^{a \dagger} = T^a,
\end{equation}
\begin{equation}
Tr \{ T^a T^b\} = \frac {1}{2} \delta^{a b},
\end{equation}
\begin{equation}
[T^a, T^b] = i f^{a b c} T^c ,
\end{equation}
\noindent whereas the field strength tensor is given by
\begin{equation}
F^a_{\mu \nu} = \partial_{\mu} A^a_{\nu} - \partial_{\nu} A^a_{\mu}
+ g f^{a b c} A^b_{\mu} A^c_{\nu}
\end{equation}
or in terms of color electric and magnetic fields as
\begin{equation}
E^a_j = i F^a_{j 0} = - \dot{A}^a_j + i \partial_j A^a_0
 + i g f^{a b c} A^b_j A^c_0 ,
\end{equation}
\begin{equation}
B^a_k = \frac{1}{2} \epsilon^{i j k } F^a_{i j } = \frac{1}{2}
\epsilon^{i j k}
(\partial_i A^a_j - \partial_j A^a_i + g f^{a b c} A^b_i A^c_j) .
\end{equation}
The Euler-Lagrange equations of motion for massless quarks then read
\begin{equation}
\gamma_{\mu} D_{\mu} \Psi = \gamma_{\mu} (\partial_{\mu} - i g T^a A^a_{\mu})
\Psi = 0
\end{equation}
and
\begin{equation}
\partial_{\mu} F^a_{\mu \nu} + g (f^{a b c} A^b_{\mu} F^c_{\mu \nu}
+ I^a_{\nu}) = 0
\end{equation}
for the gluon fields, where $I^a_{\nu}$ is the quark-color current
\begin{equation}
I^a_{\nu} = i \bar{\Psi} \gamma_{\nu} T^a \Psi .
\end{equation}

In order to perform actual calculations we have to fix the gauge. In view
of the standard Hamiltonian formulation of correlation dynamics we
choose a special axial gauge, i.e. the temporal (or Weyl) gauge \cite{Lenz}
\begin{equation}
A^a_0 = 0 ,
\end{equation}
such that the conjugate field of $A^a_i$ is (except for a sign) the
color electric field
\begin{equation}
\Pi^a_i = \dot{A}^a_i = - E^a_i .
\end{equation}
In the temporal gauge we obtain the canonical quantization conditions
\begin{equation}
[A^a_i({\bf r}, t), \Pi^b_j({\bf r'}, t)] = i \delta_{i j}
\ \delta^{a b} \ \delta^3({\bf r - r'}),
 \end{equation}
 \begin{equation}
\{\Psi({\bf r}, t), \Psi^\dagger({\bf r'}, t)\} = \delta^3({\bf r - r'}) ,
\end{equation}
while all other commutators or anticommutators vanish.
The Hamiltonian and its density then is given by ($x=({\bf r},t)\> $)
\begin{equation}
H = \int d^3r \ {\cal H}(x),
\end{equation}
with
\begin{equation}
{\cal H}(x)  = \frac{1}{2} (\Pi^a_i \Pi^a_i + B^a_i B^a_i) - g I^a_i A^a_i
+ \bar{\Psi} \gamma_i \partial_i \Psi
\end{equation}
and the Heisenberg equations of motion for $\Pi^a_i$ read
\begin{equation}
\dot{\Pi}^a_i = \frac{1}{i} [\Pi^a_i, H] = D^{a c}_k F^c_{k i} + g I^a_i
\end{equation}
with
\begin{equation}
D^{a c}_k = \delta^{a c} \partial_k + g f^{a b c} A^b_k ,
\end{equation}
while the equations of motion for $A^a_i$ reduce to (cf. (2.13))
\begin{equation}
\dot{A}^a_i = \frac{1}{i} [A^a_i, H] = \Pi^a_i .
\end{equation}
Since the Euler-Lagrange equations with $\nu = 0$ cannot be derived from the
Heisenberg equations of motion, we have to deal with additional
constraints on the gauge fields, i.e. the Gauss law  \cite{r10,Lenz}
\begin{equation}
g^a(x) = J^a(x) + \Psi^\dagger(x) T^a \Psi(x) = 0
\end{equation}
with
\begin{equation}
J^a(x) = \frac{1}{g} D^{a c}_j \Pi^c_j .
\end{equation}
In quantum physics (2.21) cannot be fulfilled as an operator
equation but has to be imposed as  constraints on the physical states
$|p\rangle$,
\begin{equation}
g^a(x)\> |p\rangle = 0.
\end{equation}
These equations are similar to Slavnov-Taylor-Ward identities within the
temporal gauge, which
impose relations between different Green functions \cite{r11}.

The basic idea of correlation dynamics now is
the use of dynamical equations of
motion for equal time Green functions, which is sufficient to describe
the time evolution of the system in a local field theory.
For this purpose, we consider the operators
$\hat{G}_n^{\lbrace a, \alpha \rbrace } (x_1, \cdots , x_n)$, which are
products of the gluon-field operator $A^a_\mu$ and its canonically conjugate
momentum $\Pi^a_\mu$ as well as quark field operators $\Psi^\dagger$ and
$\Psi$,
\begin{equation}
\hat{G}_n^{\lbrace a,\alpha \rbrace}(x_1. \cdots, x_n) =
    \hat{G}_n^{\lbrace a, \alpha \rbrace}(A^a_\mu,
\Pi^{a'}_\nu, \Psi^\dagger, \Psi),
\end{equation}
where $\lbrace a, \alpha \rbrace $ denotes a set of color and spinor
indices, respectively. $n$ is the number of field operators and all
time arguments are taken to be equal.
The equal time Green functions are given by the
quantum expectation value within the physical state $|p\rangle$
\begin{equation}
G^{\lbrace a, \alpha \rbrace}_n(x_1, \cdots, x_n) =
\langle p| \hat{G}_n^{\lbrace a, \alpha \rbrace} |p\rangle
\end{equation}
and the equations of motion for these Green functions follow from the
Heisenberg equations
\begin{equation}
\frac{d}{dt} G^{\lbrace a, \alpha\rbrace}_n(x_1, \cdots, x_n) =
 \frac{1}{i} \langle p| [\hat{G}_n^{\lbrace a, \alpha \rbrace}, H] |p\rangle .
\end{equation}
Within the canonical quantization conditions (2.14) and (2.15) equation (2.26)
leads to a coupled set of equations for the various $n$-point functions, which
for practical purposes has to be truncated.

\section{Equations of motion}

The derivation of dynamical equations of motion is straightforward, however,
somewhat lengthy. We start from the Heisenberg equations of motion (2.26)
for the quantum expectation value of the Green functions (2.25) and first
only show the lowest order equations
within the notation adopted. We use $\Psi = u^a_\beta $ with $a$ denoting
the color index and $\beta$ the spinor index, respectively. For the Dirac
matrices $\gamma_i$, $\alpha_i = \gamma_0 \gamma_i$
we use the greek indices $\alpha, \beta$
to specify the spinor components, i.e.
$\gamma^i_{\alpha \beta} = (\gamma^i)_{\alpha \beta}$,
$\alpha^i_{\alpha \beta} = (\alpha^i)_{\alpha \beta}$,
and define $t^{b a c}$ by
\begin{equation}
i t^{b a c} = (T^b)^{a c}
\end{equation}
where $b=1,...,(N^2-1)$ runs over all generators, while
$a$ and $c$ run from $1$ to $N$.
This gives for the 2-point quark Green function
\begin{eqnarray}
\label{e4}
\lefteqn{
i \frac{d}{dt} \langle \bar{u}^a_\alpha(x) u^{a'}_{\alpha'}(x')\rangle =}
 \nonumber\\
&& \alpha^i_{\beta \alpha} [\partial_{x_i}
 \langle \bar{u}^a_\beta (x) u^{a'}_{\alpha '}
(x')\rangle - g t^{b a c} \langle  \bar{u}^c_{\beta}(x) u^{a'}_{\alpha '}(x')
A^b_i(x)\rangle]
  \nonumber\\
&&+ \alpha^i_{\alpha' \beta} [\partial_{x'_i} \langle \bar{u}^a_{\alpha}(x)
u^{a'}_{\beta} (x')\rangle - g t^{b a' c} \langle \bar{u}^a_{\alpha}(x)
u^c_{\beta}(x')
A^b_i(x')\rangle] .
\end{eqnarray}
For 1- and 2-point gluon Green functions we get
\begin{equation}
\frac{d}{dt} \langle A^a_i(x)\rangle = \langle \Pi^a_i(x)\rangle ,
\end{equation}
\begin{eqnarray}
\label{e5}
\lefteqn{
\frac{d}{dt} \langle \Pi^a_i(x)\rangle = \partial^2_x \langle A^a_i(x)\rangle -
\partial_{x_i}
\partial_{x_k} \langle A^a_k(x)\rangle} \nonumber\\
&&+ 2 g f^{a b c} \partial_{y_k} \langle A^b_k(x) A^c_i(y)\rangle|_{(x=y)}
+ g^2 f^{a b c} f^{c d e} \langle A^b_k(x) A^d_k(x) A^e_i(x)\rangle
\nonumber\\
&&+ g f^{a b c} [\partial_{y_k} \langle A^b_k(y) A^c_i(x)\rangle -
\partial_{y_i}
\langle A^b_k(x) A^c_k(y)\rangle]_{(x=y)}  \nonumber\\
&&+ g t^{a b c} \gamma^i_{\alpha \beta} \langle \bar{u}^b_{\alpha}(x)
u^c_{\beta}(x)\rangle ,
\end{eqnarray}
\begin{equation}
\frac{d}{dt} \langle A^a_i(x) A^{a'}_{i'}(x')\rangle = \langle \Pi^a_i(x)
A^{a'}_{i'}(x')\rangle
+ \langle A^a_i(x) \Pi^{a'}_{i'}(x')\rangle ,
\end{equation}
\begin{eqnarray}
\label{e7}
\lefteqn{
\frac{d}{dt} \langle \Pi^a_i(x) A^{a'}_{i'}(x')\rangle = \langle \Pi^a_i(x)
\Pi^{a'}_{i'}(x')\rangle
+ \partial^2_x \langle A^a_i(x) A^{a'}_{i'}(x')\rangle}   \nonumber\\
&& - \partial_{x_i} \partial_{x_k} \langle A^a_k(x) A^{a'}_{i'}(x')\rangle
+ 2 g f^{a b c} \partial_{y_k} \langle A^b_k(x) A^c_i(y)
A^{a'}_{i'}(x')\rangle|_{(y=x)}
  \nonumber\\
&& + g^2 f^{a b c} f^{c d e}  \langle A^b_k(x) A^d_k(x) A^e_i(x)
A^{a'}_{i'}(x')\rangle
 \nonumber\\
&& + g f^{a b c} [\partial_{y_k} \langle A^b_k(y) A^c_i(x)
A^{a'}_{i'}(x')\rangle
- \partial_{y_i} \langle A^b_k(x) A^c_k(y) A^{a'}_{i'}(x')\rangle]_{(y=x)}
 \nonumber\\
&& + g t^{a b c} \gamma^i_{\alpha \beta} \langle \bar{u}^b_{\alpha}(x)
 u^c_{\beta}(x) A^{a'}_{i'}(x')\rangle
\end{eqnarray}
and
\begin{eqnarray}
\label{e6}
\lefteqn{
\frac{d}{dt} \langle \Pi^a_i(x) \Pi^{a'}_{i'}(x')\rangle = \partial^2_{x}
\langle A^a_i(x)
\Pi^{a'}_{i'}(x')\rangle + \partial^2_{x'} \langle \Pi^a_i(x)
A^{a'}_{i'}(x')\rangle} \nonumber\\
&& - \partial_{x_i} \partial_{x_k} \langle A^a_k(x) \Pi^{a'}_{i'}(x')\rangle
- \partial_{x'_i} \partial_{x'_k} \langle \Pi^a_i(x) A^{a'}_k(x')\rangle
\nonumber\\
&& + 2 g f^{a b c} \partial_{y_k} \langle A^b_k(x) A^c_i(y)
\Pi^{a'}_{i'}(x')\rangle|_{(y=x)}
\nonumber\\
&&+ 2 g f^{a' b c} \partial_{y_k} \langle \Pi^a_i(x) A^b_k(x')
A^c_{i'}(y)\rangle|_{(y=x')}
 \nonumber\\
&& + g^2 f^{a b c} f^{c d e}  \langle A^b_k(x) A^d_k(x) A^e_i(x)
\Pi^{a'}_{i'}(x')\rangle
\nonumber\\
&&+ g^2 f^{a' b c} f^{c d e} \langle \Pi^a_i(x) A^b_k(x') A^d_k(x')
A^e_{i'}(x')\rangle
 \nonumber\\
&&+ g f^{a b c} [\partial_{y_k} \langle A^b_k(y) A^c_i(x)
\Pi^{a'}_{i'}(x')\rangle
- \partial_{y_i} \langle A^b_k(x) A^c_k(y) \Pi^{a'}_{i'}(x')\rangle]_{(y=x)}
 \nonumber\\
&&+ g f^{a' b c} [ \partial_{y_k} \langle \Pi^a_i(x) A^b_k(y)
A^c_{i'}(x')\rangle
- \partial_{y_i'} \langle \Pi^a_i(x) A^b_k(x') A^c_{k}(y)\rangle]_{(y=x')}
 \nonumber\\
&& + g t^{a b c} \gamma^i_{\alpha \beta} \langle \bar{u}^b_{\alpha}(x)
 u^c_{\beta}(x) \Pi^{a'}_{i'}(x')\rangle
+ g t^{a' b c} \gamma^{i'}_{\alpha \beta} \langle \Pi^a_i(x)
\bar{u}^b_{\alpha}(x')
u^c_{\beta}(x')\rangle .
\end{eqnarray}
For the lowest order vertices one obtains
\begin{eqnarray}
\label{e8}
\lefteqn{
i \frac{d}{dt} \langle \bar{u}^a_{\alpha}(x) u^{a'}_{\alpha'}(x')
A^b_i(y)\rangle =
i \langle \bar{u}^a_{\alpha}(x) u^{a'}_{\alpha'}(x') \Pi^b_i(y)\rangle}
\nonumber\\
&& + \alpha^{k}_{\beta \alpha} [\partial_{x_k}
 \langle \bar{u}^a_{\beta}(x) u^{a'}_{\alpha'}(x') A^b_i(y)\rangle - g t^{b'a
c}
 \langle \bar{u}^c_{\beta}(x) u^{a'}_{\alpha'}(x') A^{b'}_k(x) A^b_i(y)\rangle]
 \nonumber\\
&& + \alpha^{k}_{\alpha' \beta} [\partial_{x'_k}
 \langle \bar{u}^a_{\alpha}(x) u^{a'}_{\beta}(x') A^b_i(y)\rangle - g t^{b'a'
c}
 \langle \bar{u}^a_{\alpha}(x) u^{c}_{\beta}(x') A^{b'}_k(x') A^b_i(y)\rangle]
,
\end{eqnarray}
\begin{eqnarray}
\label{e9}
\lefteqn{
i \frac{d}{dt} \langle \bar{u}^a_{\alpha}(x) u^{a'}_{\alpha'}(x')
\Pi^b_i(y)\rangle =}
\nonumber \\
&& + \alpha^{k}_{\beta \alpha} [\partial_{x_k}
 \langle \bar{u}^a_{\beta}(x) u^{a'}_{\alpha'}(x') \Pi^b_i(y)\rangle - g t^{b'a
c}
 \langle \bar{u}^c_{\beta}(x) u^{a'}_{\alpha'}(x') A^{b'}_k(x)
\Pi^b_i(y)\rangle]
 \nonumber\\
&& + \alpha^{k}_{\alpha' \beta} [\partial_{x'_k}
 \langle \bar{u}^a_{\alpha}(x) u^{a'}_{\beta}(x') \Pi^b_i(y)\rangle - g t^{b'a'
c}
 \langle \bar{u}^a_{\alpha}(x) u^{c}_{\beta}(x') A^{b'}_k(x')
\Pi^b_i(y)\rangle]
\nonumber\\
&& + i [\partial^2_y  \langle \bar{u}^a_{\alpha}(x) u^{a'}_{\alpha'}(x')
A^b_i(y)\rangle
 - \partial_{y_i} \partial_{y_k}  \langle \bar{u}^a_{\alpha}(x)
u^{a'}_{\alpha'}(x')
 A^b_k(y)\rangle   \nonumber\\
&& + 2 g f^{b b' c} \partial_{z_k}  \langle \bar{u}^a_{\alpha}(x)
u^{a'}_{\alpha'}(x')
 A^{b'}_k(y) A^c_i(z)\rangle|_{(z=y)}  \nonumber\\
&& + g^2 f^{b b' c} f^{c d e} \langle \bar{u}^a_{\alpha}(x)
u^{a'}_{\alpha'}(x')
 A^{b'}_k(y) A^d_k(y) A^e_i(y)\rangle] \nonumber\\
&& + i g f^{b b'c} [\partial_{z_k}
 \langle \bar{u}^a_{\alpha}(x) u^{a'}_{\alpha'}(x') A^{b'}_k(z) A^c_i(y)\rangle
 - \partial_{z_i}  \langle \bar{u}^a_{\alpha}(x) u^{a'}_{\alpha'}(x')
A^{b'}_k(y)
    A^c_k(z)\rangle]_{(z=y)}  \nonumber\\
&& + i g t^{b b'c} \gamma^i_{\beta\beta'}
 \langle \bar{u}^a_{\alpha}(x) u^{a'}_{\alpha'}(x')
 \bar{u}^{b'}_{\beta}(y) u^{c}_{\beta'}(y)\rangle \; .
\end{eqnarray}

In general the equations of motion for $n$-point Green functions couple to
($n$+1)- and ($n$+2)-point Green functions, as can be seen e.g. from commuting
$(\Pi ^a_i)^n$ with the cubic and quartic terms of $A^a_i$ in $H$,
such that the equations of motion
form an infinite set that has to be truncated for practical purposes.

A truncation scheme which has been quite successful in the nuclear physics
context is based on the cluster expansion of Green functions
in terms of connected (correlated) Green functions since the connected
Green functions become of minor importance with increasing order [11]. This has
been shown explicitly in the nuclear physics context in \cite{Pfitzner94}.
A nonperturbative truncation scheme that leads to a resummation of loop and
ladder diagrams in infinite order and observes the space-time conservation
laws as well as the weak Gauss law
(see next Section) is a truncation up to 4-point Green functions.

For the case of ground state Green functions the explicit expressions for
the cluster expansions can be derived from the representation of the theory
in terms of the generating functionals of full and connected
Green functions, $Z[\bar{\eta},\eta,J]$ and $W[\bar{\eta},\eta,J]$
respectively, which are given by \cite{BBJ}
\bea
Z[\bar{\eta} ,\eta, J]
=\int d\mu \left( \bar{\Psi} ,\Psi ,A_\mu \right)
exp
\left(i \int d^4x \left ({\cal L} [ \bar{\Psi} , \Psi , A_\mu ]
+ \bar{\Psi} \eta + \bar{\eta} \Psi + A_\mu J_\mu \right) \right)
\label{genfundisconn}
\eea
and
\bea
Z[\bar{\eta},\eta,J]
= exp ( W[\bar{\eta},\eta,J] ) \; ,
\label{genfunconn}
\eea
where ghosts can be omitted according to our choice of gauge.
As an example we derive the lowest order expansions in order to
present the general concept. We start with the cluster expansions for
the time-ordered Green functions with different time arguments:
\bea
\lefteqn{
\langle T \; A_i^a(x) A_{i'}^{a'}(x') \rangle
= \frac{\delta}{\delta J_i^a(x)}
\frac{\delta}{\delta J_{i'}^{a'}(x')}
exp (W[\bar{\eta},\eta,J]) |_{\bar{\eta}=\eta=J=0} } \nonumber \\
& & = \left\{
\left(
\frac{\delta}{\delta J_i^a (x)}
\frac{\delta}{\delta J_{i'}^{a'}(x')}
W[\bar{\eta},\eta,J]
\right)
exp(W[\bar{\eta},\eta,J]) \right. \nonumber \\
& & \left. + \left(
\frac{\delta}{\delta J_i^a (x)}
W[\bar{\eta},\eta,J]
\right)
\left(
\frac{\delta}{\delta J_{i'}^{a'} (x')}
W[\bar{\eta},\eta,J]
\right)
exp( W[\bar{\eta},\eta,J] )
\right\} |_{\bar{\eta}=\eta=J=0} \nonumber \\
& & = \langle T \; A_i^a(x) A_{i'}^{a'}(x') \rangle_c
+ \langle A_i^a(x) \rangle_c
\langle A_{i'}^{a'}(x') \rangle_c \; ;
\label{joernclust1}
\eea
\bea
\lefteqn{
\langle T \; \bar{u}_\alpha^a(x) u_{\alpha'}^{a'}(x') A_i^b(y) \rangle
= \frac{\delta}{\delta \eta_\alpha^a(x)}
\frac{\delta}{\delta \bar{\eta}_{\alpha'}^{a'}(x')}
\frac{1}{i}
\frac{\delta}{\delta J_i^b(y)}
exp( W[\bar{\eta},\eta,J] ) |_{\bar{\eta}=\eta=J=0} } \nonumber \\
& & = \left\{
\left(
\frac{\delta}{\delta \eta_\alpha^a(x)}
\frac{\delta}{\delta \bar{\eta}_{\alpha'}^{a'}(x')}
\frac{1}{i}
\frac{\delta}{\delta J_i^b(y)}
W[\bar{\eta},\eta,J]
\right) \right. \nonumber \\
& & + \left(
\frac{\delta}{\delta \eta_\alpha^a(x)}
W[\bar{\eta},\eta,J]
\right)
\left(
\frac{\delta}{\delta \bar{\eta}_{\alpha'}^{a'}(x')}
\frac{1}{i}
\frac{\delta}{\delta J_i^b(y)}
W[\bar{\eta},\eta,J]
\right) \nonumber \\
& & + \left(
\frac{\delta}{\delta \eta_\alpha^a(x)}
\frac{\delta}{\delta \bar{\eta}_{\alpha'}^{a'}(x')}
W[\bar{\eta},\eta,J]
\right)
\left(
\frac{1}{i}
\frac{\delta}{\delta J_i^b(y)}
W[\bar{\eta},\eta,J]
\right) \nonumber \\
& & + \left(
\frac{\delta}{\delta \eta_\alpha^a(x)}
W[\bar{\eta},\eta,J]
\right)
\left(
\frac{\delta}{\delta \bar{\eta}_{\alpha'}^{a'}}
W[\bar{\eta},\eta,J]
\right)
\left(
\frac{1}{i}
\frac{\delta}{\delta J_i^b(y)}
W[\bar{\eta},\eta,J]
\right) \nonumber \\
& & \left. + \left(
\frac{\delta}{\delta \eta_\alpha^a(x)}
\frac{1}{i}
\frac{\delta}{\delta J_i^b(y)}
W[\bar{\eta},\eta,J]
\right)
\left(
\frac{\delta}{\delta \bar{\eta}_{\alpha'}^{a'}(x')}
W[\bar{\eta},\eta,J]
\right)
\right\}
exp( W[\bar{\eta},\eta,J] ) |_{\bar{\eta}=\eta=J=0} \nonumber \\
& &=\langle T \; \bar{u}_\alpha^a(x) u_{\alpha'}^{a'}(x') A_i^b(y) \rangle_c
+ \langle T \; \bar{u}_\alpha^a(x) u_{\alpha'}^{a'}(x') \rangle_c
\langle A_i^b(y) \rangle_c \; ,
\label{joernclust2}
\eea
where all Green functions containing an unequal number of $u$ and $\bar{u}$
are assumed to vanish.
The generalization to Green functions containing conjugate gluon field
momenta is straightforward \cite{Hauser}. The expressions for equal time
ground state Green functions are obtained by taking the well-defined
equal time limit which yields the appropriate operator ordering
in the cluster expansions (\ref{joernclust1},
\ref{joernclust2}).
For the general non-equilibrium case we then simply define the connected
Green functions by using the same form of the cluster expansion as in the
ground state case.

For the lowest order Green functions one arrives at
\begin{equation}
\langle A^a_i(x) A^{a'}_{i'}(x')\rangle = \langle A^a_i(x)\rangle_c
\langle A^{a'}_{i'}(x')\rangle_c
+ \langle A^a_i(x) A^{a'}_{i'}(x')\rangle_c
\end{equation}
\begin{equation}
\langle \Pi^a_i(x) A^{a'}_{i'}(x')\rangle = \langle \Pi^a_i(x)\rangle_c
 \langle A^{a'}_{i'}(x')\rangle_c
+ \langle \Pi^a_i(x) A^{a'}_{i'}(x')\rangle_c
\end{equation}
\begin{equation}
\langle \Pi^a_i(x) \Pi^{a'}_{i'}(x')\rangle = \langle \Pi^a_i(x)\rangle_c
\langle \Pi^{a'}_{i'}(x')\rangle_c
+ \langle \Pi^a_i(x) \Pi^{a'}_{i'}(x')\rangle_c
\end{equation}
\begin{equation}
\langle \bar{u}^a_{\alpha}(x) u^{a'}_{\alpha'}(x') A^{b}_{i}(y)\rangle =
\langle \bar{u}^a_{\alpha}(x) u^{a'}_{\alpha'}(x')\rangle_c
\langle  A^{b}_{i}(y)\rangle_c +
\langle \bar{u}^a_{\alpha}(x) u^{a'}_{\alpha'}(x') A^{b}_{i}(y)\rangle_c
\end{equation}
\begin{equation}
\langle \bar{u}^a_{\alpha}(x) u^{a'}_{\alpha'}(x') \Pi^{b}_{i}(y)\rangle =
\langle \bar{u}^a_{\alpha}(x) u^{a'}_{\alpha'}(x')\rangle_c
\langle  \Pi^{b}_{i}(y)\rangle_c +
\langle \bar{u}^a_{\alpha}(x) u^{a'}_{\alpha'}(x') \Pi^{b}_{i}(y)\rangle_c ,
\end{equation}
where now the time arguments are again taken to be equal.
The explicit expressions for the relevant cluster expansions up to sixth
order are presented in Appendix A.

In reducing the dynamics to the 4-point level we assume
that all correlated terms of order $ n \geq 5$ can be neglected, i.e.
\begin{equation}
G^{\lbrace a, \alpha\rbrace}_{n c} = 0 \ {\rm for} \ \ n \geq 5 .
\end{equation}
Thus the set of equations of motion becomes closed on a finite level.

The equations of motion for the connected Green functions are obtained
by inserting the relevant cluster expansions into the equations for
the equal time Green functions ((\ref{e4})-(\ref{e9}) a.s.f.).
As an example we present here the resulting equations of motion for the
connected 1-point gluon Green functions,
\bea
\frac{d}{dt} \langle  A_{i}^{a} (x) \rangle_c = \langle  \Pi_i^a (x) \rangle_c
,
\label{gluon1a}
\eea
\bea
\lefteqn{\frac{d}{dt} \langle  \Pi_i^a (x) \rangle_c =
\partial^2_{x} \langle  A_i^a (x) \rangle_c
- \partial_{x_i} \partial_{x_k} \langle  A_k^a (x) \rangle_c}
\nn
&& + 2 g f^{abc} \partial_{y_k}[
\langle  A_k^b (x) A_i^c (y) \rangle_c + \langle  A_k^b (x) \rangle_c \langle
A_i^c (y) \rangle_c]_{(x=y)}
\nn\nn
&& + g^2 f^{abc} f^{cde} [
  \langle  A_k^b (x) A_k^d (x) A_i^e (x) \rangle_c
\nn && + \langle  A_k^b (x) \rangle_c \langle  A_k^d (x) A_i^e (x) \rangle_c
+ \langle  A_k^b (x) A_i^e (x) \rangle_c \langle  A_k^d (x) \rangle_c
\nn && + \langle  A_k^b (x) A_k^d (x) \rangle_c \langle  A_i^e (x) \rangle_c
+ \langle  A_k^b (x) \rangle_c \langle  A_k^d (x) \rangle_c \langle  A_i^e (x)
\rangle_c ]
\nn\nn
&& + g f^{abc} \partial_{y_k}
[ \langle  A_k^b (y) A_i^c (x) \rangle_c + \langle  A_k^b (y) \rangle_c \langle
 A_i^c (x) \rangle_c ]_{(x=y)}
\nn
&& - g f^{abc} \partial_{y_i}
[ \langle  A_k^b (x) A_k^c (y) \rangle_c + \langle  A_k^b (x) \rangle_c \langle
 A_k^c (y) \rangle_c ]_{(x=y)}
\nn\nn
&& + g t^{abc} \gamma_{\alpha\beta}^{i}
\langle  {\bar u}_{\alpha}^{b} (x) u_{\beta}^{c} (x) \rangle_c ,
\label{gluon1p}
\eea
\\
and the equations of motion for the connected 2-point gluon Green functions,
\bea
\frac{d}{dt} \langle  A_{i}^{a}(x) A_{i'}^{a'}(x') \rangle_c =
  \langle  \Pi_{i}^{a}(x) A_{i'}^{a'}(x') \rangle_c
+ \langle  A_{i}^{a}(x) \Pi_{i'}^{a'}(x') \rangle_c ,
\label{gluon2aa}
\eea
%
\bea
\lefteqn{\frac{d}{dt} \langle  \Pi_{i}^{a}(x) A_{i'}^{a'}(x') \rangle_c =
\langle  \Pi_{i}^{a}(x) \Pi_{i'}^{a'}(x') \rangle_c}
\nn
&&+ \partial_x^2 \langle  A_{i}^{a}(x) A_{i'}^{a'}(x') \rangle_c
- \partial_{x_i} \partial_{x_k} \langle  A_{k}^{a}(x) A_{i'}^{a'}(x') \rangle_c
\nn\nn
&&+ 2 g f^{abc} \partial_{y_k}[
  \langle  A_{k}^{b}(x) A_{i'}^{a'}(x') \rangle_c \langle  A_{i}^{c}(y)
\rangle_c
\nn
&&+ \langle  A_{k}^{b}(x) \rangle_c \langle  A_{i}^{c}(y) A_{i'}^{a'}(x')
\rangle_c
+ \langle  A_{k}^{b}(x) A_{i}^{c}(y) A_{i'}^{a'}(x') \rangle_c ]_{(x=y)}
\nn\nn
&&+ g^2 f^{abc} f^{cde} [
\langle  A_{k}^{b}(x) A_{k}^{d}(x) A_{i}^{e}(x) A_{i'}^{a'}(x') \rangle_c
\nn &&+(1 + {\cal T}_{ed} + {\cal T}_{eb}) \lbrace
\langle  A_{k}^{b}(x) \rangle_c \langle  A_{k}^{d}(x) \rangle_c \langle
A_{i}^{e}(x) A_{i'}^{a'}(x') \rangle_c
\nn && + \langle  A_{k}^{b}(x) A_{k}^{d}(x) A_{i'}^{a'}(x') \rangle_c \langle
A_{i}^{e}(x) \rangle_c
+ \langle  A_{k}^{b}(x) A_{k}^{d}(x) \rangle_c \langle  A_{i}^{e}(x)
A_{i'}^{a'}(x') \rangle_c
\rbrace ]
\nn\nn
&&+ g f^{abc} \partial_{y_k} [
  \langle  A_{k}^{b}(y) A_{i'}^{a'}(x') \rangle_c \langle  A_{i}^{c}(x)
\rangle_c
\nn
&&+ \langle  A_{k}^{b}(y) \rangle_c \langle  A_{i}^{c}(x) A_{i'}^{a'}(x')
\rangle_c
+ \langle  A_{k}^{b}(y) A_{i}^{c}(x) A_{i'}^{a'}(x') \rangle_c ]_{(x=y)}
\nn\nn
&&- g f^{abc} \partial_{y_i} [
\langle  A_{k}^{b}(x) A_{i'}^{a'}(x') \rangle_c \langle  A_{k}^{c}(y) \rangle_c
\nn
&&+\langle  A_{k}^{b}(x) \rangle_c \langle  A_{k}^{c}(y) A_{i'}^{a'}(x')
\rangle_c
+\langle  A_{k}^{b}(x) A_{k}^{c}(y) A_{i'}^{a'}(x') \rangle_c ]_{(x=y)}
\nn\nn
&&+ g t^{abc} \gamma^i_{\alpha\beta}
\langle  {\bar u}_{\alpha}^b (x) u_{\beta}^c (x) A_{i'}^{a'}(x') \rangle_c ,
\label{gluon2pa}
\eea
%
\bea
\lefteqn{\frac{d}{dt} \langle  \Pi_{i}^{a}(x) \Pi_{i'}^{a'}(x') \rangle_c =
(1+{\cal P}_{ii'}) [
\partial^2_{x}  \langle  A_{i}^{a}(x) \Pi_{i'}^{a'}(x') \rangle_c
-\partial_{x_i} \partial_{x_k} \langle  A_{k}^{a}(x) \Pi_{i'}^{a'}(x')
\rangle_c ]}
\nn\nn
&&+2g (1+{\cal P}_{ii'}) f^{abc} \partial_{y_k} [
\langle  A_{k}^{b}(x) \Pi_{i'}^{a'}(x') \rangle_c \langle  A_{i}^{c}(y)
\rangle_c
\nn
&&+ \langle  A_{k}^{b}(x) \rangle_c \langle  A_{i}^{c}(y) \Pi_{i'}^{a'}(x')
\rangle_c
+ \langle  A_{k}^{b}(x) A_{i}^{c}(y) \Pi_{i'}^{a'}(x') \rangle_c ]_{(x=y)}
\nn\nn
&&+ g^2 (1+{\cal P}_{ii'}) f^{abc} f^{cde}[
\langle  A_{k}^{b}(x) A_{k}^{d}(x) A_{i}^{e}(x) \Pi_{i'}^{a'}(x') \rangle_c
\nn &&+ (1 + {\cal T}_{ed} + {\cal T}_{eb}) \lbrace
\langle  A_{k}^{b}(x) \rangle_c \langle  A_{k}^{d}(x) \rangle_c \langle
A_{i}^{e}(x) \Pi_{i'}^{a'}(x') \rangle_c
\nn &&+ \langle  A_{k}^{b}(x) A_{k}^{d}(x) \Pi_{i'}^{a'}(x') \rangle_c \langle
A_{i}^{e}(x) \rangle_c
+ \langle  A_{k}^{b}(x) A_{k}^{d}(x) \rangle_c \langle  A_{i}^{e}(x)
\Pi_{i'}^{a'}(x') \rangle_c
\rbrace ]
\nn\nn
&&+g (1+{\cal P}_{ii'}) f^{abc} \partial_{y_k}[
\langle  A_{k}^{b}(y) \Pi_{i'}^{a'}(x') \rangle_c \langle  A_{i}^{c}(x)
\rangle_c
\nn
&&+\langle  A_{k}^{b}(y) \rangle_c \langle  A_{i}^{c}(x) \Pi_{i'}^{a'}(x')
\rangle_c
+ \langle  A_{k}^{b}(y) A_{i}^{c}(x) \Pi_{i'}^{a'}(x') \rangle_c ]_{(x=y)}
\nn\nn
&&-g (1+{\cal P}_{ii'}) f^{abc} \partial_{y_i}[
\langle  A_{k}^{b}(x) \Pi_{i'}^{a'}(x') \rangle_c \langle  A_{k}^{c}(y)
\rangle_c
\nn
&&+\langle  A_{k}^{b}(x) \rangle_c \langle  A_{k}^{c}(y) \Pi_{i'}^{a'}(x')
\rangle_c
+\langle  A_{k}^{b}(x) A_{k}^{c}(y) \Pi_{i'}^{a'}(x') \rangle_c ]_{ (x=y)}
\nn\nn
&&+g (1+{\cal P}_{ii'}) t^{abc} \gamma^i_{\alpha\beta}
\langle  {\bar u}_{\alpha}^b (x) u_{\beta}^c (x) \Pi_{i'}^{a'}(x') \rangle_c.
\label{gluon2pp}
\eea
In  (\ref{gluon2pa}) and (\ref{gluon2pp}) we have introduced
two kinds of permutation operators ${\cal P}_{ii'}$ and ${\cal T}_{ed}$
in order to achieve an unambiguous compactification.
To demonstrate the action of these operators we look  as an example
at parts of the $g^2$-terms in (\ref{gluon2pp})
\bea
\lefteqn{
g^2 (1+{\cal P}_{ii'}) f^{abc} f^{cde} (1+{\cal T}_{ed}+{\cal T}_{eb})
\langle  A_{k}^{b}(x) \rangle_c \langle  A_{k}^{d}(x) \rangle_c
\langle  A_{i}^{e}(x) \Pi_{i'}^{a'}(x') \rangle_c}
\nn\nn &&
= g^2 (1+{\cal P}_{ii'}) f^{abc} f^{cde} [
\langle  A_{k}^{b}(x) \rangle_c \langle  A_{k}^{d}(x) \rangle_c
\langle  A_{i}^{e}(x) \Pi_{i'}^{a'}(x') \rangle_c
\nn &&
+\langle  A_{k}^{b}(x) \rangle_c \langle  A_{i}^{e}(x) \rangle_c
\langle  A_{k}^{d}(x) \Pi_{i'}^{a'}(x') \rangle_c
+\langle  A_{i}^{e}(x) \rangle_c \langle  A_{k}^{d}(x) \rangle_c
\langle  A_{k}^{b}(x) \Pi_{i'}^{a'}(x') \rangle_c ]
\nn\nn &&
= g^2 f^{abc} f^{cde} [
\langle  A_{k}^{b}(x) \rangle_c \langle  A_{k}^{d}(x) \rangle_c
\langle  A_{i}^{e}(x) \Pi_{i'}^{a'}(x') \rangle_c
\nn &&
+\langle  A_{k}^{b}(x) \rangle_c \langle  A_{i}^{e}(x) \rangle_c
\langle  A_{k}^{d}(x) \Pi_{i'}^{a'}(x') \rangle_c
+\langle  A_{i}^{e}(x) \rangle_c \langle  A_{k}^{d}(x) \rangle_c
\langle  A_{k}^{b}(x) \Pi_{i'}^{a'}(x') \rangle_c ]
\nn &&
+ g^2 f^{a'bc} f^{cde} [
\langle  A_{k}^{b}(x') \rangle_c \langle  A_{k}^{d}(x') \rangle_c
\langle  \Pi_{i}^{a}(x)  A_{i'}^{e}(x') \rangle_c
\nn &&
+\langle  A_{k}^{b}(x') \rangle_c \langle  A_{i'}^{e}(x') \rangle_c
\langle  \Pi_{i}^{a}(x) A_{k}^{d}(x') \rangle_c
+\langle  A_{i'}^{e}(x') \rangle_c \langle  A_{k}^{d}(x') \rangle_c
\langle  \Pi_{i}^{a}(x) A_{k}^{b}(x') \rangle_c ].
\label{example}
\eea
Obviously ${\cal P}_{ii'}$ and ${\cal T}_{ed}$ act in the same way
by interchanging the fields or conjugate momenta labeled by
the spatial components $i,i'$ (in case of ${\cal P}_{ii'}$)
or by the color indices $e,d$ (in case of ${\cal T}_{ed}$).
Due to their length we have shifted the remaining equations of motion for the
connected 3- and 4-point gluon Green functions to Appendix B.

For the connected 2- and 4-point quark Green functions we obtain
\bea
\lefteqn{
i \frac{d}{dt} \langle \ubr_{\alpha}^{a} (x) u_{\alpha'}^{a'} (x') \rangle_c =}
\nn &&
\alpha^{i}_{\beta\alpha} [ \partial_{x_i}  \langle  \ubr _{\beta}^{a}(x)  u
_{\alpha'}^{a'}(x')  \rangle_c
- g  t^{bac} \lbrace
  \langle  \ubr _{\beta}^{c}(x)  u  _{\alpha'}^{a'}(x') A _{i}^{b}(x)
\rangle_c
+ \langle  \ubr _{\beta}^{c}(x)  u  _{\alpha'}^{a'}(x')  \rangle_c
  \langle  A _{i}^{b}(x)  \rangle_c \rbrace ]
\nn &&
+  \alpha^{i}_{\alpha'\beta} [
\partial_{x'_i}  \langle  \ubr _{\alpha}^{a}(x)  u  _{\beta}^{a'}(x')
\rangle_c
- g  t^{ba'c} \lbrace
  \langle  \ubr _{\alpha}^{a}(x)  u  _{\beta}^{c}(x')  A _{i}^{b}(x')
\rangle_c
+ \langle  \ubr _{\alpha}^{a}(x)  u  _{\beta}^{c}(x')  \rangle_c
\langle  A _{i}^{b}(x')  \rangle_c \rbrace ]
\nn
\label{quark2}
\eea
and
\bea
\lefteqn{
i \frac{d}{dt} \langle \ubr_{\alpha'}^{a'} (x') \ubr_{\alpha''}^{a''} (x'')
 u_{\beta''}^{b''} (y'') u_{\beta'}^{b'} (y') \rangle_c =
\alpha^{i}_{\beta\alpha'} [
\partial_{x'_i}  \langle  \ubr _{\beta}^{a'}(x')  \ubr _{\alpha''}^{a''}(x'')
u  _{\beta''}^{b''}(y'')  u  _{\beta'}^{b'}(y')  \rangle_c}
\nn &&
- g  t^{ba'c}  \lbrace
  \langle  \ubr _{\beta}^{c}(x')  \ubr _{\alpha''}^{a''}(x'')  u
_{\beta''}^{b''}(y'')  u  _{\beta'}^{b'}(y')  \rangle_c  \langle  A
_{i}^{b}(x')  \rangle_c
+ \langle  \ubr _{\beta}^{c}(x')  u  _{\beta'}^{b'}(y')  \rangle_c  \langle
\ubr _{\alpha''}^{a''}(x'')  u  _{\beta''}^{b''}(y'')  A _{i}^{b}(x')
\rangle_c
\nn &&
- \langle  \ubr _{\beta}^{c}(x')  u  _{\beta''}^{b''}(y'')  \rangle_c  \langle
\ubr _{\alpha''}^{a''}(x'')  u  _{\beta'}^{b'}(y')  A _{i}^{b}(x')  \rangle_c
\rbrace]
\nn\nn &&
+  \alpha^{i}_{\beta'\beta} [
\partial_{y'_i}  \langle  \ubr _{\alpha'}^{a'}(x')  \ubr _{\alpha''}^{a''}(x'')
 u  _{\beta''}^{b''}(y'')  u  _{\beta}^{b'}(y')  \rangle_c
- g  t^{bb'c}  \lbrace
  \langle  \ubr _{\alpha'}^{a'}(x')  \ubr _{\alpha''}^{a''}(x'')  u
_{\beta''}^{b''}(y'')  u  _{\beta}^{c}(y')  \rangle_c  \langle  A _{i}^{b}(y')
\rangle_c
\nn &&
- \langle  \ubr _{\alpha'}^{a'}(x')  u  _{\beta''}^{b''}(y'')  A _{i}^{b}(y')
\rangle_c  \langle  \ubr _{\alpha''}^{a''}(x'')  u  _{\beta}^{c}(y')  \rangle_c
+ \langle  \ubr _{\alpha'}^{a'}(x')  u  _{\beta}^{c}(y')  \rangle_c  \langle
\ubr _{\alpha''}^{a''}(x'')  u  _{\beta''}^{b''}(y'')  A _{i}^{b}(y')
\rangle_c  \rbrace ]
\nn\nn &&
+  \alpha^{i}_{\beta\alpha''}  [
\partial_{x''_i}  \langle  \ubr _{\alpha'}^{a'}(x')  \ubr _{\beta}^{a''}(x'')
u  _{\beta''}^{b''}(y'')  u  _{\beta'}^{b'}(y')  \rangle_c
- g  t^{ba''c}  \lbrace
  \langle  \ubr _{\alpha'}^{a'}(x')  \ubr _{\beta}^{c}(x'')  u
_{\beta''}^{b''}(y'')  u  _{\beta'}^{b'}(y')  \rangle_c  \langle  A
_{i}^{b}(x'')  \rangle_c
\nn &&
+ \langle  \ubr _{\alpha'}^{a'}(x')  u  _{\beta'}^{b'}(y')  A _{i}^{b}(x'')
\rangle_c  \langle  \ubr _{\beta}^{c}(x'')  u  _{\beta''}^{b''}(y'')  \rangle_c
- \langle  \ubr _{\alpha'}^{a'}(x')  u  _{\beta''}^{b''}(y'')  A _{i}^{b}(x'')
\rangle_c  \langle  \ubr _{\beta}^{c}(x'')  u  _{\beta'}^{b'}(y')  \rangle_c
\rbrace ]
\nn\nn &&
+  \alpha^{i}_{\beta''\beta} [\partial_{y''_i}  \langle  \ubr
_{\alpha'}^{a'}(x')  \ubr _{\alpha''}^{a''}(x'')  u  _{\beta}^{b''}(y'')  u
_{\beta'}^{b'}(y')  \rangle_c
- g  t^{bb''c}  \lbrace
  \langle  \ubr _{\alpha'}^{a'}(x')  \ubr _{\alpha''}^{a''}(x'')  u
_{\beta}^{c}(y'')  u  _{\beta'}^{b'}(y')  \rangle_c  \langle  A _{i}^{b}(y'')
\rangle_c
\nn &&
+ \langle  \ubr _{\alpha'}^{a'}(x')  u  _{\beta'}^{b'}(y')  A _{i}^{b}(y'')
\rangle_c  \langle  \ubr _{\alpha''}^{a''}(x'')  u  _{\beta}^{c}(y'')
\rangle_c
- \langle  \ubr _{\alpha'}^{a'}(x')  u  _{\beta}^{c}(y'')  \rangle_c  \langle
\ubr _{\alpha''}^{a''}(x'')  u  _{\beta'}^{b'}(y')  A _{i}^{b}(y'')  \rangle_c
\rbrace ] ,
\nn
\label{quark4}
\eea
which represents all equations on the pure quark sector.

The equations of motion for the connected 3-point quark-gluon vertex
Green functions are given by
\bea
\lefteqn{
i \frac{d}{dt} \langle \ubr_{\alpha}^{a} (x) u_{\alpha'}^{a'} (x') A_{i}^{b}(y)
\rangle_c =
i \langle  \ubr _{\alpha}^{a}(x)  u  _{\alpha'}^{a'}(x')  \Pi _{i}^{b}(y)
\rangle_c }
\nn &&
+  \alpha^{k}_{\beta\alpha}  [
\partial_{x_k}  \langle  \ubr _{\beta}^{a}(x)  u  _{\alpha'}^{a'}(x')  A
_{i}^{b}(y)  \rangle_c
- g  t^{b_0 a c}  \lbrace
\langle  \ubr _{\beta}^{c}(x)  u  _{\alpha'}^{a'}(x')  A _{k}^{b_0}(x)  A
_{i}^{b}(y)  \rangle_c
\nn &&
+ \langle  \ubr _{\beta}^{c}(x)  u  _{\alpha'}^{a'}(x')  \rangle_c  \langle  A
_{k}^{b_0}(x)  A _{i}^{b}(y)  \rangle_c
+ \langle  \ubr _{\beta}^{c}(x)  u  _{\alpha'}^{a'}(x')  A _{i}^{b}(y)
\rangle_c  \langle  A _{k}^{b_0}(x)  \rangle_c \rbrace]
\nn\nn &&
+  \alpha^{k}_{\alpha'\beta} [
\partial_{x'_k}  \langle  \ubr _{\alpha}^{a}(x)  u  _{\beta}^{a'}(x')  A
_{i}^{b}(y)  \rangle_c
- g  t^{b_0 a'c}  \lbrace
  \langle  \ubr _{\alpha}^{a}(x)  u  _{\beta}^{c}(x')  A _{k}^{b_0}(x')  A
_{i}^{b}(y)  \rangle_c
\nn &&
+ \langle  \ubr _{\alpha}^{a}(x)  u  _{\beta}^{c}(x')  \rangle_c  \langle  A
_{k}^{b_0}(x')  A _{i}^{b}(y)  \rangle_c
+ \langle  \ubr _{\alpha}^{a}(x)  u  _{\beta}^{c}(x')  A _{i}^{b}(y)  \rangle_c
 \langle  A _{k}^{b_0}(x')  \rangle_c \rbrace]
\label{mixed3a}
\eea
and
%
\bea
\lefteqn{
i \frac{d}{dt} \langle \ubr_{\alpha}^{a} (x) u_{\alpha'}^{a'} (x')
\Pi_{i}^{b}(y) \rangle_c = \alpha^{k}_{\beta\alpha}  [
\partial_{x_k}  \langle  \ubr _{\beta}^{a}(x)  u  _{\alpha'}^{a'}(x')  \Pi
_{i}^{b}(y)  \rangle_c }
\nn &&
-g  t^{b_0 a c}  \lbrace
  \langle  \ubr _{\beta}^{c}(x)  u  _{\alpha'}^{a'}(x')  A _{k}^{b_0}(x)  \Pi
_{i}^{b}(y)  \rangle_c
+ \langle  \ubr _{\beta}^{c}(x)  u  _{\alpha'}^{a'}(x')  \rangle_c  \langle  A
_{k}^{b_0}(x)  \Pi _{i}^{b}(y)  \rangle_c
\nn &&
+ \langle  \ubr _{\beta}^{c}(x)  u  _{\alpha'}^{a'}(x')  \Pi _{i}^{b}(y)
\rangle_c  \langle  A _{k}^{b_0}(x)  \rangle_c \rbrace]
\nn\nn &&
+  \alpha^{k}_{\alpha'\beta}  [
\partial_{x'_k}  \langle  \ubr _{\alpha}^{a}(x)  u  _{\beta}^{a'}(x')  \Pi
_{i}^{b}(y)  \rangle_c
- g  t^{b_0 a'c}  \lbrace
  \langle  \ubr _{\alpha}^{a}(x)  u  _{\beta}^{c}(x')  A _{k}^{b_0}(x')  \Pi
_{i}^{b}(y)  \rangle_c
\nn &&
+ \langle  \ubr _{\alpha}^{a}(x)  u  _{\beta}^{c}(x')  \rangle_c  \langle  A
_{k}^{b_0}(x')  \Pi _{i}^{b}(y)  \rangle_c
+ \langle  \ubr _{\alpha}^{a}(x)  u  _{\beta}^{c}(x')  \Pi _{i}^{b}(y)
\rangle_c  \langle  A _{k}^{b_0}(x')  \rangle_c \rbrace]
\nn\nn &&
+ i  \partial^2_y  \langle  \ubr _{\alpha}^{a}(x)  u  _{\alpha'}^{a'}(x')  A
_{i}^{b}(y)  \rangle_c
- i  \partial_{y_i}  \partial_{y_k}  \langle  \ubr _{\alpha}^{a}(x)  u
_{\alpha'}^{a'}(x')  A _{k}^{b}(y)  \rangle_c
\nn\nn &&
+ i  2 g  f^{b b_0 c}  \partial_{z_k}  [
   \langle  \ubr _{\alpha}^{a}(x)  u  _{\alpha'}^{a'}(x')  A _{k}^{b_0}(y)  A
_{i}^{c}(z)  \rangle_c
\nn &&
+  \langle  \ubr _{\alpha}^{a}(x)  u  _{\alpha'}^{a'}(x')  A _{k}^{b_0}(y)
\rangle_c  \langle  A _{i}^{c}(z)  \rangle_c
+  \langle  \ubr _{\alpha}^{a}(x)  u  _{\alpha'}^{a'}(x')  A _{i}^{c}(z)
\rangle_c  \langle  A _{k}^{b_0}(y)  \rangle_c  ]_{(y=z)}
\nn\nn &&
+ i g^2  f^{b b_0 c}  f^{cde}  (1 + {\cal T}_{b_0 d} + {\cal T}_{b_0 e})[
   \langle  \ubr _{\alpha}^{a}(x)  u  _{\alpha'}^{a'}(x')  A _{k}^{b_0}(y)
\rangle_c  \langle  A _{k}^{d}(y)  A _{i}^{e}(y)  \rangle_c
\nn &&
+  \langle  \ubr _{\alpha}^{a}(x)  u  _{\alpha'}^{a'}(x')  A _{k}^{b_0}(y)
\rangle_c  \langle  A _{k}^{d}(y)  \rangle_c  \langle  A _{i}^{e}(y)  \rangle_c
+  \langle  \ubr _{\alpha}^{a}(x)  u  _{\alpha'}^{a'}(x')  A _{k}^{d}(y)  A
_{i}^{e}(y)  \rangle_c  \langle  A _{k}^{b_0}(y)  \rangle_c ]
\nn\nn &&
+ ig  f^{b b_0 c}  \partial_{z_k}  [
   \langle  \ubr _{\alpha}^{a}(x)  u  _{\alpha'}^{a'}(x')  A _{k}^{b_0}(z)  A
_{i}^{c}(y)  \rangle_c
\nn &&
+  \langle  \ubr _{\alpha}^{a}(x)  u  _{\alpha'}^{a'}(x')  A _{k}^{b_0}(z)
\rangle_c  \langle  A _{i}^{c}(y)  \rangle_c
+  \langle  \ubr _{\alpha}^{a}(x)  u  _{\alpha'}^{a'}(x')  A _{i}^{c}(y)
\rangle_c  \langle  A _{k}^{b_0}(z)  \rangle_c  ]_{(y=z)}
\nn\nn &&
- ig  f^{b b_0 c}  \partial_{z_i} [
   \langle  \ubr _{\alpha}^{a}(x)  u  _{\alpha'}^{a'}(x')  A _{k}^{b_0}(y)  A
_{k}^{c}(z)  \rangle_c
\nn &&
+  \langle  \ubr _{\alpha}^{a}(x)  u  _{\alpha'}^{a'}(x')  A _{k}^{b_0}(y)
\rangle_c  \langle  A _{k}^{c}(z)  \rangle_c
+  \langle  \ubr _{\alpha}^{a}(x)  u  _{\alpha'}^{a'}(x')  A _{k}^{c}(z)
\rangle_c  \langle  A _{k}^{b_0}(y)  \rangle_c  ]_{(y=z)}
\nn\nn &&
+ ig  t^{ba'c}  \gamma ^{i}_{\alpha'\beta'}  \delta (x'- y)  \langle  \ubr
_{\alpha}^{a}(x)  u  _{\beta'}^{c}(y)  \rangle_c
\nn &&
+ ig  t^{b b_0 c}  \gamma ^{i}_{\beta\beta'}  [
  \langle  \ubr _{\alpha}^{a}(x)  \ubr _{\beta}^{b_0}(y)  u  _{\beta'}^{c}(y)
u  _{\alpha'}^{a'}(x')  \rangle_c
- \langle  \ubr _{\alpha}^{a}(x)  u  _{\beta'}^{c}(y)  \rangle_c  \langle  \ubr
_{\beta}^{b_0}(y)  u  _{\alpha'}^{a'}(x')  \rangle_c  ],
\label{mixed3b}
\eea
while the explicit equations of motion for the connected 4-point
mixed quark and gluon
Green functions are shifted to Appendix B again.

The closed set of equations (\ref{gluon1a}) - (3.24) and (3.26) - (3.29)
including (\ref{gluon3aaa}) to (\ref{mixed4pp}) will be denoted
as {\bf C}onstraint {\bf QCD} ({\bf CQCD}) equations furtheron.

\section{Compatibility with Gauss law and Ward \protect\\ identities}

The aim of the present section is to show the compatibility of the
{\bf CQCD} equations with the Gauss law and the Ward
identities, respectively (cf. \cite{Wang93}).
For this purpose, we first note
that the operators $g^a(x)$
at equal times constitute a set of local generators for the SU(N) gauge
group because of \cite{r10}
\begin{equation}
[g^a({\bf r}; t), g^b({\bf r'}, t)] = i f^{a b c} g^c({\bf r}, t)
\ \delta^3({\bf r - r'}) .
\label{4.1}
\end{equation}
Furthermore, the Gauss law operator commutes with the Hamiltonian \cite{r10}
\begin{equation}
[g^a({\bf r}, t), H] = 0
\label{4.2}
\end{equation}
which expresses the conservation of $g^a$ in time
\begin{equation}
\frac{d}{dt} g^a({\bf r},t) = \frac{1}{i} [g^a({\bf r},t), H] = 0.
\label{4.3}
\end{equation}
Equations (\ref{4.2}) and (\ref{4.3}) imply that within the temporal gauge
the system evolved in time by $H$ has a residual gauge symmetry and the
Gauss operators $g^a({\bf r},t)$ are the local generators of the residual gauge
symmetry.

Now we assume the actual realization of the Gauss law as a quantum
expectation value
\begin{equation}
\langle p|g^a({\bf r},t)|p\rangle = 0
\label{4.4}
\end{equation}
instead of (2.23). This may be considered as a weak form of the Gauss law,
since it is less restrictive for the physical states than (2.23). For example,
the perturbative vacuum $|0\rangle$ fulfills (\ref{4.4}) but not (2.23)
\cite{r12}. Also
(\ref{4.4}) allows for the local propagation of colored objects, whereas
(2.23) restricts to color-singlet objects.
These consequences of (\ref{4.4}) are exactly
what we aim at for describing properties of the quark-gluon plasma.

Furthermore, since all $g^a(x)$ are conserved
quantities and the field equations of motion are generated by $H$, the
weak Gauss law is conserved identically in time provided that it is fulfilled
initially. This is just what we have been aiming at: The Hamiltonian
dynamics are compatible with the conservation of the weak Gauss law, and the
problem is shifted to an initial value problem.
This statement still holds within 4-point correlation dynamics, since
the Gauss law operator contains two-point operators at most.

Whereas the weak Gauss law is a constraint imposed on the lowest order Green
functions of the fields by the residual gauge symmetry, the Ward identities
impose constraints on higher order Green functions due to the same residual
gauge symmetry.

We start defining a Lie operation $L_{g^a({\bf r},t)}$ by
\begin{equation}
L_{g^a({\bf r},t)} g^b({\bf r'},t) =: [g^a({\bf r},t), g^b({\bf r'},t)].
\end{equation}
{}From the algebraic structure of $g^a({\bf r},t)$ (\ref{4.1}) we get
\begin{equation}
L_{g^{a_1}({\bf r_1}, t)} L_{g^{a_2}({\bf r_2},t)} g^{a_3}({\bf r_3}, t) =
(i)^2 \ f^{a_1 b c} f^{a_2 a_3 b} \delta^3({\bf r_2 - r_3})
\delta^3({\bf r_1 - r_2}) g^c({\bf r_1}, t),
\end{equation}
or
\begin{equation}
L_{g^{a_1}({\bf r_1}, t)} L_{g^{a_2}({\bf r_2},t)}
\cdots L_{g^{a_{n-1}}({\bf r_{n-1}}, t)}  g^{a_n}({\bf r_n}, t) =
F^{a_1 \cdots a_n a_{n+1}}({\bf r_1, \cdots, r_n})
 g^{a_{n+1}}({\bf r_1}, t),
\end{equation}
where $F^{a_1 \cdots a_n a_{n+1}}({\bf r_1, \cdots, r_n})$ is a
complex function of
$\{{\bf r_1, \cdots, r_n}\}$.
Thus we obtain from (\ref{4.4})
\begin{equation}
\langle p| [ g^{a_1} ({\bf r_1},t),
g^{a_2} ({\bf r_2},t) ] |p \rangle = 0
\label{4.8}
\end{equation}
\bea
\langle p|L_{g^{a_1}({\bf r_1}, t)} L_{g^{a_2}({\bf r_2},t)}
 g^{a_3}({\bf r_3}, t)|p\rangle = 0
\label{4.9}
\eea
and
\begin{equation}
\langle p|L_{g^{a_1}({\bf r_1}, t)} L_{g^{a_2}({\bf r_2},t)}
\cdots L_{g^{a_{n-1}}({\bf r_{n-1}}, t)}  g^{a_n}({\bf r_n}, t)|p\rangle = 0 .
\label{4.10}
\end{equation}

The relation between (\ref{4.8}), (\ref{4.9}), (\ref{4.10}) and gauge
invariance can be established
as follows. Since $H$ due to the temporal gauge has a residual gauge
symmetry generated by the local algebra
$\{g^a({\bf r},t)\}$, the residual group $U(g)$ will generate
a set ${\cal H}_{g p}$ of physical states $|p\rangle_g$ from any physical state
$|p\rangle$
\begin{equation}
|p\rangle_g = U(g) |p\rangle; \  {\rm for} \ |p\rangle_g \epsilon \ {\cal H}_{p
g} ,
\end{equation}
where the residual group element is given by
\begin{equation}
U(g) =  \exp\left (i \int d^3r\> g^b \omega^b \right ),
\end{equation}
where $\omega^b=\omega^b({\bf r})$ are the local gauge transformation
angles.
The state $|p\rangle_g$ has the same energy as $|p\rangle$ since
\begin{equation}
_g\langle p|H|p\rangle_g = \langle p|U^{-1}(g) H U(g)|p\rangle = \langle p| H
|p\rangle .
\end{equation}
The weak Gauss law is fulfilled as well
\begin{eqnarray}
_g\langle p| g^a |p\rangle_g & = & \langle p|  \exp(-i \int g^b \omega^b \
d^3r)\>
g^a\> \exp(i \int g^c \omega^c \ d^3r) |p\rangle  \nonumber\\
& = & \langle p| U_{a b}\> g^b |p\rangle  =  0,
\label{zang}
\end{eqnarray}
where $U_{a b}$ is a color matrix containing the gauge transformation angles.

{}From the theory of Lie groups it is evident that (\ref{4.10}) and
(\ref{zang})
are equivalent. Furthermore, since $U(g)$ generates a gauge transformation
for the SU(N) gauge fields, any gauge invariant physical observable $O$ has
the same expectation value within the set ${\cal H}_{p g}$.

Since the weak Gauss law is conserved in time within a 4-point truncation
scheme,
all Ward identities which follow from the group properties (\ref{4.1}) of the
Gauss law operators in combination with the weak Gauss law,
such as (\ref{4.8}), (\ref{4.9}) and (\ref{4.10}), will also be conserved.
This is due to the fact that correlation dynamics in general conserves the
commutation and anticommutation relations of all field operators within Green
functions, which leads to a conservation of the expectation values of all
operator equations generated by these relations.
The problem of the Ward identities thus is again shifted to an
initial value problem.

\section{Summary and discussion}

In this work we have derived a closed set of equations of motion for
connected Green functions of quark and SU(N) gluon fields up to the
4-point level. The resulting set of equations, which we for abbreviation
will denote by {\bf C}onstraint {\bf QCD} ({\bf CQCD}) equations furtheron,
fulfill the conservation of linear momentum, angular-momentum and total
energy throughout time as well as the
weak Gauss law and the Ward identities - as generated by commutators of
Gauss operators - provided that the initial conditions
follow the gauge constraints. The latter may be achieved,
for example, by a functional containing
the chromomagnetic field (2.8), which obeys the Gauss
law \cite{r13}. It might be used as a trial wave function in a variational
calculation within a finite number of basis states. Alternatively, one might
start from perturbative gluon-field configurations and increase the
coupling $g$ from zero adiabatically for the preparation of an initial
state \footnote{Note that the perturbative vacuum $|0\rangle$ is a
physical state in our
realization of the gauge constraint because it fulfills (4.4).}.

We note that the CQCD equations can easily be transformed into
transport equations by means of a Wigner transformation with respect to
the space-time degrees of freedom, which is nothing but a unitary
transformation.
Since the latter transformation has been performed quite often in the
literature, we do not present the final expressions for reasons of missing
compactness.

The actual goal is a numerical integration of the CQCD equations within a
finite
basis set in analogy to the works performed in the nuclear physics context
\cite{T27} - \cite{P93}
which appear to quite naturally explain a variety of non-perturbative nuclear
phenomena \cite{C23}. Though the actual expressions in Appendix B appear very
cumbersome, we note that the reduction of the quark-gluon many-body problem
-- as e.g. studied by lattice QCD in thermal equilibrium --
to a nonlinear two-body problem provides a numerical task of lower
complexity than the original QCD equations on the lattice. Furthermore,
the CQCD equations allow to study the dynamical evolution of non-equilibrium
configurations of quarks and gluons and to explore the
low-energy QCD dynamics, especially the linear response of systems close
to the ground state. However, it is not clear if the actual solutions of our
approach will reproduce all the physical phenomena of the full QCD equations.
Before speculating about the convergence of the CQCD equations we prefer to
analyse the numerical results first. Such work is in progress.

\vspace{1cm}
One of the authors (S. J. W.) likes to thank Prof. U. Mosel for the kind
hospitality at the University of Giessen where this work was performed.


\newpage
\section*{Appendix}
\zeqn
\begin{appendix}
\section{Cluster expansions}
By taking the functional derivative of the generating functional
(\ref{genfundisconn}) with
respect to the different sources (cf. Sect. 3) we obtain the
relevant cluster expansions of Green functions up to sixth order
in the gluon, quark, and mixed sector of the SU(N) gauge theory.
For reasons of compactness we introduce the general particle
exchange operators
\bea
{\cal S}_{l}^{k} =
( 1 + \sum_{n=l}^{k} [ {\cal P}_{1n} + {\cal P}_{2n} ]
    + \sum_{n=l}^{k-1} \sum_{j=n+1}^{k} {\cal P}_{1n} {\cal P}_{2j} ),
\label{a.1}
\eea
where ${\cal P}_{ij}$ is the two-body permutation operator.
Here $i$ and $j$ represent the set of
quantum numbers characterizing the gauge fields, the corresponding conjugate
momenta or the quark spinors.

We note that any application of
the ${\cal P}_{ij}$ operators has to ensure the original
order of the field operators within the connected Green functions.
\subsection{Gluon sector}

In the following the integer numbers between brackets stand for gauge fields
$A^a_i$ or conjugate momenta $\Pi^a_i$.
The cluster expansions then can be written in compact form as:
\bea
\langle 1 \rangle = \langle 1 \rangle_c ,
\hspace*{13cm}
\label{a.2}
\eea
\bea
\langle 12 \rangle = \langle 12 \rangle_c
+ \langle 1 \rangle\langle 2 \rangle,
\hspace*{11cm}
\label{a.3}
\eea
\bea
\langle 123 \rangle=\langle 123 \rangle_c
+(1+ \sum_{n=2}^{3} {\cal P}_{1n})
\langle 1 \rangle\langle 23 \rangle_c
+\langle 1 \rangle\langle 2 \rangle\langle 3 \rangle,
\hspace*{6cm}
\label{a.4}
\eea
\bea
\lefteqn{\langle 1234 \rangle = \langle 1234 \rangle_c
+ (1 +\sum_{n=2}^{4} {\cal P}_{1n}) \langle 1 \rangle\langle 234 \rangle_c
+ ( 1+ \sum_{n=3}^{4} {\cal P}_{2n})
\langle 12 \rangle_c \langle 34 \rangle_c }
\nn
&&+ {\cal S}_{3}^{4}\langle 1 \rangle\langle 2 \rangle\langle 34 \rangle_c
+ \langle 1 \rangle\langle 2 \rangle\langle 3 \rangle\langle 4 \rangle,
\hspace*{9cm}
\label{a.5}
\eea
\bea
\lefteqn{\langle 12345 \rangle
= (1+ \sum_{n=2}^{5} {\cal P}_{1n}) \langle 1 \rangle\langle 2345 \rangle_c}
\nn
&& + {\cal S}_{3}^{5}
   [ \langle 12\rangle_c \langle 345\rangle_c
+ \langle 1\rangle\langle 2\rangle\langle 345\rangle_c
+ \langle 12\rangle_c \langle 3\rangle\langle 4\rangle\langle 5\rangle ]
\nn
&& + ( 1+ \sum_{n=1}^{4} {\cal P}_{5n} )( 1 + \sum_{n=3}^{4} {\cal P}_{2n} )
\langle 12 \rangle_c \langle 34 \rangle_c \langle 5 \rangle
+\langle 1\rangle\langle 2\rangle\langle 3\rangle\langle 4\rangle\langle
5\rangle,
\hspace*{4cm}
\label{a.6}
\eea
\bea
\lefteqn{\langle 123456\rangle =
(1 + \sum_{n=4}^{6} \sum_{j=1}^{3}{\cal P}_{jn})
[ \langle 123\rangle_c\langle 456\rangle_c
+ ( 1 + {\cal P}_{14} {\cal P}_{25} {\cal P}_{36} )
\langle 123 \rangle_c \langle 4 \rangle \langle 5 \rangle \langle 6 \rangle ]}
\nn
&& + {\cal S}_{3}^{6}
[ \langle 12\rangle _c \langle 3456\rangle_c
+ \langle 1\rangle\langle 2\rangle\langle 3456\rangle_c
+ \langle 12\rangle_c\langle 3\rangle
  \langle 4\rangle\langle 5\rangle\langle 6\rangle ]
\nn
&& + {\cal S}_{3}^{6} (1+ \sum_{n=5}^{6}{\cal P}_{4n})
\langle 1 \rangle \langle 2 \rangle \langle 34 \rangle_c \langle 56\rangle_c
 + (1+ \sum_{n=1}^{5} {\cal P}_{6n}) {\cal S}_{3}^{5}
\langle 12 \rangle_c \langle 345 \rangle_c \langle 6 \rangle
\nn
&& + ( 1 + \sum_{n=1}^{4} {\cal P}_{5n}) ( 1 + \sum_{n=3}^{4} {\cal P}_{2n})
\langle 12\rangle_c\langle 34\rangle_c\langle 56\rangle_c
+ \langle 1\rangle\langle 2\rangle\langle 3\rangle
\langle 4\rangle\langle 5\rangle\langle 6\rangle.
\hspace*{2cm}
\label{a.7}
\eea
The number of terms for the 1-, 2-,..., 5-, and 6-point gluon Green
functions consist out of 1, 2, 5, 15, 52, and 203 different contributions,
respectively.
Due to the applied truncation scheme we have neglected the connected
5-point function $\langle 12345 \rangle_c$ in  (\ref{a.6})
and the terms $(1+ \sum_{n=2}^{6}{\cal P}_{1n}) \langle 1\rangle
\langle 23456\rangle_c$ and $\langle 123456\rangle_c$
in (\ref{a.7}).
\subsection{Quark sector}
The argument $i$ of a quark spinor $u(i)$ in this subsection stands for
the spatial coordinates and the Dirac, flavor and color indices of a quark $i$.
The cluster expansions then can be written as:
\bea
\langle \ubr(1) u(1')\rangle = \langle \ubr(1) u(1')\rangle_c ,
\hspace*{10cm}
\label{a.8}
\eea
\bea
\langle \ubr(1) \ubr(2) u(2') u(1') \rangle =
\langle \ubr(1) \ubr(2) u(2') u(1') \rangle_c
+ (1- {\cal P}_{1'2'})
\langle \ubr(1) u(1')\rangle_c \langle \ubr(2) u(2')\rangle_c,
\nn
\label{a.9}
\eea
\bea
\lefteqn{\langle \ubr(1) \ubr(2)\ubr(3) u(3') u(2') u(1') \rangle = }
\nn && (1 -{\cal P}_{1'2'} - {\cal P}_{1'3'} -{\cal P}_{2'3'}
+{\cal P}_{23}( {\cal P}_{1'3'} + {\cal P}_{1'2'}))
\langle \ubr(1) u(1')\rangle_c \langle \ubr(2) u(2')\rangle_c \langle \ubr(3)
u(3')\rangle_c
\nn
&&+ (1 - {\cal P}_{1'2'} -{\cal P}_{1'3'})
\langle \ubr(1) u(1')\rangle_c \langle \ubr(2)\ubr(3) u(3') u(2')\rangle_c
\nn
&&+ (1 - {\cal P}_{1'2'} -{\cal P}_{2'3'})
\langle \ubr(2) u(2')\rangle_c \langle \ubr(1)\ubr(3) u(3') u(1')\rangle_c
\nn
&&+ (1 - {\cal P}_{1'3'} -{\cal P}_{2'3'})
\langle \ubr(3) u(3')\rangle_c \langle \ubr(1)\ubr(2) u(2') u(1')\rangle_c.
\label{a.10}
\eea
The number of terms for the 2-, 4-, and 6-point quark Green functions
consist out of 1, 3, and 16 contributions, respectively.
Due to the truncation on the 4-point level we have neglected
the connected 6-point function
$\langle \ubr(1)\ubr(2)\ubr(3) u(3') u(2') u(1') \rangle_c$
in (\ref{a.10}).
\subsection{Mixed sector}
In the mixed sector of the SU(N) gauge theory the Green functions
contain quark and gluon field operators simultaneously. Here
the $u(i)$ represent the quark spinors while the numbers 2, 3, 4, and 5
stand for gauge fields or conjugate momenta as in Subsection A.1.
Since the explicit cluster expansions in the gluon sector are presented
in (\ref{a.2}) - (\ref{a.7}) we have not specified the expansions of
the pure gluonic Green functions in (\ref{a.11}) - (\ref{a.16}).
The expansions up to the 4-point connected level read:
\bea
\langle  \ubr(1) u(1')2\rangle = \langle  \ubr(1) u(1')2 \rangle_c
+ \langle  \ubr(1) u(1')\rangle_c \langle 2\rangle,
\hspace*{5cm}
\label{a.11}
\eea
\bea
\langle \ubr(1) u(1') 23\rangle =
\langle \ubr(1) u(1') 23\rangle_c
+ \langle \ubr(1) u(1')\rangle_c \langle 23\rangle
+ ( 1+ {\cal P}_{23}) \langle \ubr(1) u(1') 2\rangle_c \langle 3\rangle,
\label{a.12}
\eea
\bea
\lefteqn{\langle \ubr(1) u(1') 234\rangle =
\langle \ubr(1) u(1')\rangle_c \langle 234\rangle
+ ( 1+ {\cal P}_{23} + {\cal P}_{24})
\langle \ubr(1) u(1')2\rangle_c \langle 34\rangle}
\nn
&& + ( 1+ {\cal P}_{24} + {\cal P}_{34})
\langle \ubr(1) u(1')23 \rangle_c \langle 4\rangle,
\hspace*{8cm}
\label{a.13}
\eea
\bea
\lefteqn{ \langle \ubr(1) u(1') 2345 \rangle =
\langle \ubr(1) u(1')\rangle_c \langle 2345\rangle
+ ( 1+ \sum_{n=3}^{5} {\cal P}_{2n})
\langle \ubr(1) u(1')2\rangle_c \langle 345\rangle}
\nn
&& + ( 1+ \sum_{n=4}^{5} [ {\cal P}_{2n} + {\cal P}_{3n}]
   + {\cal P}_{24}{\cal P}_{35})
\langle \ubr(1) u(1')23 \rangle_c \langle 45\rangle,
\hspace*{4cm}
\label{a.14}
\eea
\bea
\lefteqn{ \langle \ubr(1) \ubr(2) u(2') u(1')3 \rangle =
\langle \ubr(1) \ubr(2) u(2') u(1')\rangle_c \langle 3\rangle}
\nn
&& + (1- {\cal P}_{1'2'})
[\langle \ubr(1) u(1')\rangle_c \langle \ubr(2)u(2')\rangle_c \langle 3\rangle
+\langle \ubr(1) u(1')3\rangle_c \langle \ubr(2) u(2')\rangle_c
\nn &&+\langle \ubr(1) u(1')\rangle_c \langle \ubr(2) u(2')3\rangle_c ] ,
\label{a.15}
\eea
\bea
\lefteqn{\langle \ubr(1)\ubr(2) u(2')u(1')34\rangle =
\langle \ubr(1) \ubr(2) u(2')u(1')\rangle_c \langle 34\rangle}
\nn
&& + ( 1- {\cal P}_{1'2'})
[ \langle \ubr(1) u(1') \rangle_c \langle \ubr(2) u(2')\rangle_c \langle
34\rangle
+ \langle \ubr(1) u(1') 34\rangle_c \langle \ubr(2) u(2')\rangle_c
\nn
&& + \langle \ubr(1) u(1')\rangle_c \langle \ubr(2) u(2')34\rangle_c
+ (1+{\cal P}_{34})
\langle  \ubr(1) u(1') 3\rangle_c \langle \ubr(2) u(2') 4\rangle_c ].
\label{a.16}
\eea
The original cluster expansions (\ref{a.11}) - (\ref{a.16}) consist
out of 2, 5, 15, 52, 8, and 17 terms.
Because of the truncation scheme on the 4-point level we have neglected
$\langle \ubr(1) u(1') 234 \rangle_c$ in (\ref{a.13}),
$ \langle \ubr(1) u(1') 2345 \rangle_c$ and
$ (1+ \sum_{n=3}^{5}{\cal P}_{2n})
\langle \ubr(1) u(1') 345 \rangle_c \langle 2 \rangle$ in

\noindent
(\ref{a.14}),
$ \langle \ubr(1) \ubr(2) u(2') u(1') 3 \rangle_c$ in (\ref{a.15}),
and $\langle \ubr(1)\ubr(2) u(2') u(1') 34 \rangle_c$ and

\noindent
$ (1+ {\cal P}_{34})
\langle \ubr(1)\ubr(2) u(2') u(1') 3 \rangle_c \langle 4\rangle $
in (\ref{a.16}).
%
\section{Equations of motion for connected Green functions}
\subsection{Gluon sector}
Whereas the time evolution of the 1-point and 2-point gluon Green functions
is given in (\ref{gluon1a}) - (\ref{gluon2pp}) we here continue with
the 3- and 4-point functions.
In order to allow for a compact presentation we again use
the permutation operators ${\cal P}_{ii'}$ and ${\cal T}_{ed}$
as defined in Section 3.
The equations of motion for the connected 3-point gluon Green functions read:
%
\bea
\lefteqn{
\frac{d}{dt} \langle  A_{i}^{a}(x) A_{i'}^{a'}(x') A_{i''}^{a''} (x'')
\rangle_c =
 \langle  \Pi_{i}^{a}(x) A_{i'}^{a'}(x') A_{i''}^{a''}(x'') \rangle_c}
\nn
&&+ \langle  A_{i}^{a}(x) \Pi_{i'}^{a'}(x') A_{i''}^{a''}(x'') \rangle_c
+ \langle  A_{i}^{a}(x) A_{i'}^{a'}(x') \Pi_{i''}^{a''}(x'') \rangle_c ,
\label{gluon3aaa}
\eea
%
\bea
\lefteqn{
\frac{d}{dt} \langle  \Pi_{i}^{a}(x) A_{i'}^{a'}(x') A_{i''}^{a''} (x'')
\rangle_c =
\langle  \Pi_{i}^{a}(x) \Pi_{i'}^{a'}(x') A_{i''}^{a''}(x'') \rangle_c
+ \langle  \Pi_{i}^{a}(x) A_{i'}^{a'}(x') \Pi_{i''}^{a''}(x'') \rangle_c}
\nn\nn
&&+ \partial^2_{x} \langle  A_{i}^{a}(x) A_{i'}^{a'}(x') A_{i''}^{a''}(x'')
\rangle_c
- \partial_{x_i} \partial_{x_k}  \langle  A_{k}^{a}(x) A_{i'}^{a'}(x')
A_{i''}^{a''}(x'') \rangle_c
\nn\nn
&&+ 2 g f^{abc} \partial_{y_k} [
\langle  A_{k}^{b}(x) A_{i}^{c}(y) A_{i'}^{a'}(x') A_{i''}^{a''}(x'') \rangle_c
\nn &&+(1+ {\cal T}_{cb}) \lbrace
\langle  A_{k}^{b}(x) A_{i'}^{a'}(x') A_{i''}^{a''}(x'') \rangle_c \langle
A_{i}^{c}(y) \rangle_c
\nn &&+(1+ {\cal T}_{a'a''})
\langle  A_{k}^{b}(x) A_{i'}^{a'}(x') \rangle_c \langle  A_{i}^{c}(y)
A_{i''}^{a''}(x'') \rangle_c
\rbrace ]_{(x=y)}
\nn\nn
&&+ g^2 f^{abc} f^{cde}
(1+ {\cal T}_{eb} + {\cal T}_{ed}) [
\langle  A_{k}^{b}(x) A_{k}^{d}(x) A_{i'}^{a'}(x') A_{i''}^{a''}(x'') \rangle_c
\langle  A_{i}^{e}(x) \rangle_c
\nn &&+(1+{\cal T}_{bd}) \langle  A_{k}^{b}(x) A_{i'}^{a'}(x') \rangle_c
\langle  A_{k}^{d}(x) A_{i''}^{a''}(x'') \rangle_c \langle  A_{i}^{e}(x)
\rangle_c
\nn &&+\langle  A_{k}^{b}(x) \rangle_c \langle  A_{k}^{d}(x) \rangle_c \langle
A_{i}^{e}(x) A_{i'}^{a'}(x') A_{i''}^{a''}(x'') \rangle_c
\nn &&+(1+{\cal T}_{a'a''})\langle  A_{k}^{b}(x) A_{k}^{d}(x) A_{i'}^{a'}(x')
\rangle_c \langle  A_{i}^{e}(x) A_{i''}^{a''}(x'') \rangle_c
\nn &&+\langle  A_{k}^{b}(x) A_{k}^{d}(x) \rangle_c \langle  A_{i}^{e}(x)
A_{i'}^{a'}(x') A_{i''}^{a''}(x'') \rangle_c]
\nn\nn
&&+ g f^{abc} \partial_{y_k} [
\langle  A_{k}^{b}(y) A_{i}^{c}(x) A_{i'}^{a'}(x') A_{i''}^{a''}(x'') \rangle_c
\nn && + (1+{\cal T}_{cb}) \lbrace
\langle  A_{k}^{b}(y) A_{i'}^{a'}(x') A_{i''}^{a''}(x'') \rangle_c \langle
A_{i}^{c}(x) \rangle_c
\nn && + (1+ {\cal T}_{a'a''})
\langle  A_{k}^{b}(y) A_{i'}^{a'}(x') \rangle_c \langle  A_{i}^{c}(x)
A_{i''}^{a''}(x'') \rangle_c
\rbrace ]_{(x=y)}
\nn\nn
&&- g f^{abc} \partial_{y_i} [
\langle  A_{k}^{b}(x) A_{k}^{c}(y) A_{i'}^{a'}(x') A_{i''}^{a''}(x'') \rangle_c
\nn && + (1+ {\cal T}_{cb})
\langle  A_{k}^{b}(x) A_{i'}^{a'}(x') A_{i''}^{a''}(x'') \rangle_c \langle
A_{k}^{c}(y) \rangle_c
\nn&& + (1+ {\cal T}_{a'a''})
\langle  A_{k}^{b}(x) A_{i'}^{a'}(x') \rangle_c \langle  A_{k}^{c}(y)
A_{i''}^{a''}(x'') \rangle_c
\rbrace ]_{(x=y)}
\nn\nn
&&+ g t^{abc} \gamma_{\alpha\beta}^{i}
\langle  {\bar u}_{\alpha}^b (x) u_{\beta}^c (x) A_{i'}^{a'}(x')
A_{i''}^{a''}(x'') \rangle_c ,
\label{gluon3paa}
\eea
%
\bea
\lefteqn{
\frac{d}{dt} \langle  \Pi_{i}^{a}(x) \Pi_{i'}^{a'}(x') A_{i''}^{a''} (x'')
\rangle_c =
\langle  \Pi_{i}^{a}(x) \Pi_{i'}^{a'}(x') \Pi_{i''}^{a''}(x'') \rangle_c }
\nn\nn
&&+(1+{\cal P}_{ii'})
[\partial^2_{x} \langle  A_{i}^{a}(x) \Pi_{i'}^{a'}(x') A_{i''}^{a''}(x'')
\rangle_c
- \partial_{x_i} \partial_{x_k}  \langle  A_{k}^{a}(x) \Pi_{i'}^{a'}(x')
A_{i''}^{a''}(x'') \rangle_c ]
\nn\nn
&&+ 2 g (1+{\cal P}_{ii'})
f^{abc} \partial_{y_k} [
\langle  A_{k}^{b}(x) A_{i}^{c}(y) \Pi_{i'}^{a'}(x') A_{i''}^{a''}(x'')
\rangle_c
\nn && + (1+ {\cal T}_{cb}) \lbrace
\langle  A_{k}^{b}(x) \Pi_{i'}^{a'}(x') A_{i''}^{a''}(x'') \rangle_c \langle
A_{i}^{c}(y) \rangle_c
\nn&& + (1+ {\cal T}_{a'a''})
 \langle  A_{k}^{b}(x) \Pi_{i'}^{a'}(x') \rangle_c \langle  A_{i}^{c}(y)
A_{i''}^{a''}(x'') \rangle_c
\rbrace ]_{(x=y)}
\nn\nn
&&+ g^2 (1+{\cal P}_{ii'}) f^{abc} f^{cde}
(1+ {\cal T}_{eb} + {\cal T}_{ed})
\nn &&\times [\langle  A_{k}^{b}(x) A_{k}^{d}(x) \Pi_{i'}^{a'}(x')
A_{i''}^{a''}(x'') \rangle_c \langle  A_{i}^{e}(x) \rangle_c
\nn &&+(1+{\cal T}_{bd}) \langle  A_{k}^{b}(x) \Pi_{i'}^{a'}(x') \rangle_c
\langle  A_{k}^{d}(x) A_{i''}^{a''}(x'') \rangle_c \langle  A_{i}^{e}(x)
\rangle_c
\nn &&+\langle  A_{k}^{b}(x) \rangle_c \langle  A_{k}^{d}(x) \rangle_c \langle
A_{i}^{e}(x) \Pi_{i'}^{a'}(x') A_{i''}^{a''}(x'') \rangle_c
\nn &&+(1+{\cal T}_{a'a''})\langle  A_{k}^{b}(x) A_{k}^{d}(x) \Pi_{i'}^{a'}(x')
\rangle_c \langle  A_{i}^{e}(x) A_{i''}^{a''}(x'') \rangle_c
\nn &&+\langle  A_{k}^{b}(x) A_{k}^{d}(x) \rangle_c \langle  A_{i}^{e}(x)
\Pi_{i'}^{a'}(x') A_{i''}^{a''}(x'') \rangle_c]
\nn\nn
&&+ g (1+{\cal P}_{ii'})
f^{abc} \partial_{y_k} [
\langle  A_{k}^{b}(y) A_{i}^{c}(x) \Pi_{i'}^{a'}(x') A_{i''}^{a''}(x'')
\rangle_c
\nn &&+ (1+ {\cal T}_{cb}) \lbrace
\langle  A_{k}^{b}(y) \Pi_{i'}^{a'}(x') A_{i''}^{a''}(x'') \rangle_c \langle
A_{i}^{c}(x) \rangle_c
\nn &&+ (1+ {\cal T}_{a'a''})
 \langle  A_{k}^{b}(y) \Pi_{i'}^{a'}(x') \rangle_c \langle  A_{i}^{c}(x)
A_{i''}^{a''}(x'') \rangle_c
\rbrace ]_{(x=y)}
\nn\nn
&&- g (1+{\cal P}_{ii'})
f^{abc} \partial_{y_i} [
\langle  A_{k}^{b}(x) A_{k}^{c}(y) \Pi_{i'}^{a'}(x') A_{i''}^{a''}(x'')
\rangle_c
\nn &&+ (1+ {\cal T}_{cb}) \lbrace
\langle  A_{k}^{b}(x) \Pi_{i'}^{a'}(x') A_{i''}^{a''}(x'') \rangle_c \langle
A_{k}^{c}(y) \rangle_c
\nn &&+ (1+ {\cal T}_{a'a''})
\langle  A_{k}^{b}(x) \Pi_{i'}^{a'}(x') \rangle_c \langle  A_{k}^{c}(y)
A_{i''}^{a''}(x'') \rangle_c
\rbrace ]_{(x=y)}
\nn\nn
&&+ g (1+{\cal P}_{ii'})
t^{abc} \gamma_{\alpha\beta}^{i}
\langle  {\bar u}_{\alpha}^b (x) u_{\beta}^c (x)  \Pi_{i'}^{a'}(x')
A_{i''}^{a''}(x'') \rangle_c ,
\label{gluon3ppa}
\eea
%
\bea
\lefteqn{
\frac{d}{dt} \langle  \Pi_{i}^{a}(x) \Pi_{i'}^{a'}(x') \Pi_{i''}^{a''} (x'')
\rangle_c =}
\nn
&&(1 + {\cal P}_{ii'} + {\cal P}_{ii''})
[ \partial^2_{x} \langle  A_{i}^{a}(x) \Pi_{i'}^{a'}(x') \Pi_{i''}^{a''}(x'')
\rangle_c
- \partial_{x_i}  \partial_{x_k} \langle  A_{k}^{a}(x) \Pi_{i'}^{a'}(x')
\Pi_{i''}^{a''}(x'') \rangle_c ]
\nn\nn
&&+ 2 g (1 + {\cal P}_{ii'} + {\cal P}_{ii''})
f^{abc} \partial_{y_k} [
\langle  A_{k}^{b}(x) A_{i}^{c}(y) \Pi_{i'}^{a'}(x') \Pi_{i''}^{a''}(x'')
\rangle_c
\nn && + (1+ {\cal T}_{cb}) \lbrace
\langle  A_{k}^{b}(x) \Pi_{i'}^{a'}(x') \Pi_{i''}^{a''}(x'') \rangle_c \langle
A_{i}^{c}(y) \rangle_c
\nn && +(1+ {\cal T}_{a'a''})
 \langle  A_{k}^{b}(x) \Pi_{i'}^{a'}(x') \rangle_c \langle  A_{i}^{c}(y)
\Pi_{i''}^{a''}(x'') \rangle_c
\rbrace ]_{(x=y)}
\nn\nn
&&+ g^2 (1 + {\cal P}_{ii'} + {\cal P}_{ii''}) f^{abc} f^{cde}
(1+ {\cal T}_{eb} + {\cal T}_{ed})
\nn &&\times [\langle  A_{k}^{b}(x) A_{k}^{d}(x) \Pi_{i'}^{a'}(x')
\Pi_{i''}^{a''}(x'') \rangle_c \langle  A_{i}^{e}(x) \rangle_c
\nn &&+(1+{\cal T}_{bd}) \langle  A_{k}^{b}(x) \Pi_{i'}^{a'}(x') \rangle_c
\langle  A_{k}^{d}(x) \Pi_{i''}^{a''}(x'') \rangle_c \langle  A_{i}^{e}(x)
\rangle_c
\nn &&+\langle  A_{k}^{b}(x) \rangle_c \langle  A_{k}^{d}(x) \rangle_c \langle
A_{i}^{e}(x) \Pi_{i'}^{a'}(x') \Pi_{i''}^{a''}(x'') \rangle_c
\nn &&+(1+{\cal T}_{a'a''})\langle  A_{k}^{b}(x) A_{k}^{d}(x) \Pi_{i'}^{a'}(x')
\rangle_c \langle  A_{i}^{e}(x) A_{i''}^{a''}(x'') \rangle_c
\nn &&+\langle  A_{k}^{b}(x) A_{k}^{d}(x) \rangle_c \langle  A_{i}^{e}(x)
\Pi_{i'}^{a'}(x') \Pi_{i''}^{a''}(x'') \rangle_c]
\nn\nn
&&+ g (1 + {\cal P}_{ii'} + {\cal P}_{ii''})
f^{abc} \partial_{y_k} [
\langle  A_{k}^{b}(y) A_{i}^{c}(x) \Pi_{i'}^{a'}(x') \Pi_{i''}^{a''}(x'')
\rangle_c
\nn && + (1+ {\cal T}_{cb}) \lbrace
\langle  A_{k}^{b}(y) \Pi_{i'}^{a'}(x') \Pi_{i''}^{a''}(x'') \rangle_c \langle
A_{i}^{c}(x) \rangle_c
\nn && + (1+ {\cal T}_{a'a''})
\langle  A_{k}^{b}(y) \Pi_{i'}^{a'}(x') \rangle_c \langle  A_{i}^{c}(x)
\Pi_{i''}^{a''}(x'') \rangle_c
\rbrace ]_{(x=y)}
\nn\nn
&&- g (1 + {\cal P}_{ii'} + {\cal P}_{ii''})
f^{abc} \partial_{y_i} [
\langle  A_{k}^{b}(x) A_{k}^{c}(y) \Pi_{i'}^{a'}(x') \Pi_{i''}^{a''}(x'')
\rangle_c
\nn && + (1+ {\cal T}_{cb}) \lbrace
\langle  A_{k}^{b}(x) \Pi_{i'}^{a'}(x') \Pi_{i''}^{a''}(x'') \rangle_c \langle
A_{k}^{c}(y) \rangle_c
\nn && + (1+ {\cal T}_{a'a''})
\langle  A_{k}^{b}(x) \Pi_{i'}^{a'}(x') \rangle_c \langle  A_{k}^{c}(y)
\Pi_{i''}^{a''}(x'') \rangle_c
\rbrace ]_{(x=y)}
\nn\nn
&&+ g (1 + {\cal P}_{ii'} + {\cal P}_{ii''})
t^{abc} \gamma_{\alpha\beta}^{i}
\langle  {\bar u}_{\alpha}^b (x)  u_{\beta}^c(x) \Pi_{i'}^{a'}(x')
\Pi_{i''}^{a''}(x'') \rangle_c.
\label{gluon3ppp}
\eea
\\
The equations of motion for connected 4-point gluon Green functions
are the most cumbersome of our approach and read:
%
\bea
\lefteqn{
\frac{d}{dt} \langle  A_{i}^{a}(x) A_{i'}^{a'}(x') A_{i''}^{a''} (x'')
A_{i'''}^{a'''} (x''') \rangle_c =
 \langle  \Pi_{i}^{a}(x) A_{i'}^{a'}(x') A_{i''}^{a''}(x'')
A_{i'''}^{a'''}(x''') \rangle_c}
\nn
&&+ \langle  A_{i}^{a}(x) \Pi_{i'}^{a'}(x') A_{i''}^{a''}(x'')
A_{i'''}^{a'''}(x''') \rangle_c
+ \langle  A_{i}^{a}(x) A_{i'}^{a'}(x') \Pi_{i''}^{a''}(x'')
A_{i'''}^{a'''}(x''') \rangle_c
\nn
&&+ \langle  A_{i}^{a}(x) A_{i'}^{a'}(x') A_{i''}^{a''}(x'')
\Pi_{i'''}^{a'''}(x''') \rangle_c,
\label{gluon4aaaa}
\eea
%
\bea
\lefteqn{
\frac{d}{dt} \langle  \Pi_{i}^{a}(x) A_{i'}^{a'}(x') A_{i''}^{a''} (x'')
A_{i'''}^{a'''} (x''') \rangle_c =
\langle  \Pi_{i}^{a}(x) \Pi_{i'}^{a'}(x') A_{i''}^{a''}(x'')
A_{i'''}^{a'''}(x''') \rangle_c}
\nn
&&+ \langle  \Pi_{i}^{a}(x) A_{i'}^{a'}(x') \Pi_{i''}^{a''}(x'')
A_{i'''}^{a'''}(x''') \rangle_c
+ \langle  \Pi_{i}^{a}(x) A_{i'}^{a'}(x') A_{i''}^{a''}(x'')
\Pi_{i'''}^{a'''}(x''') \rangle_c
\nn\nn
&&+ \partial^2_{x} \langle  A_{i}^{a}(x) A_{i'}^{a'}(x') A_{i''}^{a''}(x'')
A_{i'''}^{a'''}(x''') \rangle_c
- \partial_{x_i}  \partial_{x_k} \langle  A_{k}^{a}(x) A_{i'}^{a'}(x')
A_{i''}^{a''}(x'') A_{i'''}^{a'''}(x''') \rangle_c
\nn\nn
&&+ 2 g f^{abc} \partial_{y_k} (1+ {\cal T}_{cb})[
\langle  A_{k}^{b}(x) A_{i'}^{a'}(x') A_{i''}^{a''}(x'') A_{i'''}^{a'''}(x''')
\rangle_c \langle  A_{i}^{c}(y) \rangle_c
\nn&&+(1+ {\cal T}_{a'a'''} +{\cal T}_{a''a'''})
\langle  A_{k}^{b}(x) A_{i'}^{a'}(x') A_{i''}^{a''}(x'') \rangle_c \langle
A_{i}^{c}(y) A_{i'''}^{a'''}(x''') \rangle_c]_{(x=y)}
\nn\nn
&&+ g^2 f^{abc} f^{cde}
(1+ {\cal T}_{eb} + {\cal T}_{ed})
\nn &&\times [
( \langle  A_{k}^{b}(x) \rangle_c \langle  A_{k}^{d}(x) \rangle_c + \langle
A_{k}^{b}(x) A_{k}^{d}(x) \rangle_c )
\langle  A_{i}^{e}(x) A_{i'}^{a'}(x') A_{i''}^{a''}(x'') A_{i'''}^{a'''}(x''')
\rangle_c
\nn &&+(1+{\cal T}_{db}) \lbrace
\langle  A_{k}^{b}(x) A_{i'}^{a'}(x') \rangle_c \langle  A_{k}^{d}(x)
A_{i''}^{a''}(x'') \rangle_c \langle  A_{i}^{e}(x) A_{i'''}^{a'''}(x''')
\rangle_c
\nn &&+(1+{\cal T}_{a'a'''} + {\cal T}_{a''a'''})
\langle  A_{k}^{b}(x) A_{i'}^{a'}(x') A_{i''}^{a''}(x'') \rangle_c
\langle  A_{k}^{d}(x) A_{i'''}^{a'''}(x''') \rangle_c
\langle  A_{i}^{e}(x) \rangle_c
\rbrace
\nn &&+(1+{\cal T}_{a'a''} + {\cal T}_{a'a'''})
\lbrace
\langle  A_{k}^{b}(x)  A_{k}^{d}(x) A_{i'}^{a'}(x') \rangle_c
\langle  A_{i}^{e}(x) A_{i''}^{a''}(x'') A_{i'''}^{a'''}(x''') \rangle_c
\nn &&+
\langle  A_{k}^{b}(x)  A_{k}^{d}(x) A_{i''}^{a''}(x'') A_{i'''}^{a'''}(x''')
\rangle_c
\langle  A_{i}^{e}(x) A_{i'}^{a'}(x') \rangle_c
\rbrace ]
\nn\nn
&&+ g f^{abc} \partial_{y_k}(1+ {\cal T}_{cb})[
\langle  A_{k}^{b}(y) A_{i'}^{a'}(x') A_{i''}^{a''}(x'') A_{i'''}^{a'''}(x''')
\rangle_c \langle  A_{i}^{c}(x) \rangle_c
\nn&&+(1+ {\cal T}_{a'a'''} + {\cal T}_{a''a'''})
\langle  A_{k}^{b}(y) A_{i'}^{a'}(x') A_{i''}^{a''}(x'') \rangle_c \langle
A_{i}^{c}(x) A_{i'''}^{a'''}(x''') \rangle_c]_{(x=y)}
\nn\nn
&&- g f^{abc} \partial_{y_i}(1+ {\cal T}_{cb})[
\langle  A_{k}^{b}(x) A_{i'}^{a'}(x') A_{i''}^{a''}(x'') A_{i'''}^{a'''}(x''')
\rangle_c \langle  A_{k}^{c}(y) \rangle_c
\nn&&+(1+ {\cal T}_{a'a'''} + {\cal T}_{a''a'''})
\langle  A_{k}^{b}(x) A_{i'}^{a'}(x') A_{i''}^{a''}(x'') \rangle_c \langle
A_{k}^{c}(y) A_{i'''}^{a'''}(x''') \rangle_c]_{(x=y)} ,
\label{gluon4paaa}
\eea
%
\bea
\lefteqn{
\frac{d}{dt} \langle  \Pi_{i}^{a}(x) \Pi_{i'}^{a'}(x') A_{i''}^{a''} (x'')
A_{i'''}^{a'''} (x''') \rangle_c =
 \langle  \Pi_{i}^{a}(x) \Pi_{i'}^{a'}(x') \Pi_{i''}^{a''}(x'')
A_{i'''}^{a'''}(x''') \rangle_c}
\nn
&&+ \langle  \Pi_{i}^{a}(x) \Pi_{i'}^{a'}(x') A_{i''}^{a''}(x'')
\Pi_{i'''}^{a'''}(x''') \rangle_c
\nn\nn
&&+ (1+ {\cal P}_{ii'}) [
\partial^2_{x}  \langle  A_{i}^{a}(x) \Pi_{i'}^{a'}(x') A_{i''}^{a''}(x'')
A_{i'''}^{a'''}(x''') \rangle_c
\nn && - \partial_{x_i} \partial_{x_k} \langle  A_{k}^{a}(x) \Pi_{i'}^{a'}(x')
A_{i''}^{a''}(x'') A_{i'''}^{a'''}(x''') \rangle_c]
\nn\nn
&&+ 2 g (1+ {\cal P}_{ii'})
f^{abc} \partial_{y_k} (1+ {\cal T}_{cb})[
\langle  A_{k}^{b}(x) \Pi_{i'}^{a'}(x') A_{i''}^{a''}(x'')
A_{i'''}^{a'''}(x''') \rangle_c \langle  A_{i}^{c}(y) \rangle_c
\nn&&+(1+ {\cal T}_{a'a'''} + {\cal T}_{a''a'''})
\langle  A_{k}^{b}(x) \Pi_{i'}^{a'}(x') A_{i''}^{a''}(x'') \rangle_c \langle
A_{i}^{c}(y) A_{i'''}^{a'''}(x''') \rangle_c]_{(x=y)}
\nn\nn
&&+ g^2 (1+ {\cal P}_{ii'})
f^{abc} f^{cde} (1+ {\cal T}_{eb} + {\cal T}_{ed})
\nn &&\times [
( \langle  A_{k}^{b}(x) \rangle_c \langle  A_{k}^{d}(x) \rangle_c + \langle
A_{k}^{b}(x) A_{k}^{d}(x) \rangle_c )
\langle  A_{i}^{e}(x) \Pi_{i'}^{a'}(x') A_{i''}^{a''}(x'')
A_{i'''}^{a'''}(x''') \rangle_c
\nn &&+(1+{\cal T}_{db}) \lbrace
\langle  A_{k}^{b}(x) \Pi_{i'}^{a'}(x') \rangle_c \langle  A_{k}^{d}(x)
A_{i''}^{a''}(x'') \rangle_c \langle  A_{i}^{e}(x) A_{i'''}^{a'''}(x''')
\rangle_c
\nn &&+(1+{\cal T}_{a'a'''} + {\cal T}_{a''a'''})
\langle  A_{k}^{b}(x) \Pi_{i'}^{a'}(x') A_{i''}^{a''}(x'') \rangle_c
\langle  A_{k}^{d}(x) A_{i'''}^{a'''}(x''') \rangle_c
\langle  A_{i}^{e}(x) \rangle_c
\rbrace
\nn &&+(1+{\cal T}_{a'a''} + {\cal T}_{a'a'''})
\lbrace
\langle  A_{k}^{b}(x)  A_{k}^{d}(x) \Pi_{i'}^{a'}(x') \rangle_c
\langle  A_{i}^{e}(x) A_{i''}^{a''}(x'') A_{i'''}^{a'''}(x''') \rangle_c
\nn &&+
\langle  A_{k}^{b}(x)  A_{k}^{d}(x) A_{i''}^{a''}(x'') A_{i'''}^{a'''}(x''')
\rangle_c
\langle  A_{i}^{e}(x) \Pi_{i'}^{a'}(x') \rangle_c
\rbrace ]
\nn\nn
&&+ g (1+ {\cal P}_{ii'})
f^{abc} \partial_{y_k}(1+ {\cal T}_{cb}) [
\langle  A_{k}^{b}(y) \Pi_{i'}^{a'}(x') A_{i''}^{a''}(x'')
A_{i'''}^{a'''}(x''') \rangle_c \langle  A_{i}^{c}(x) \rangle_c
\nn&&+(1+ {\cal T}_{a'a'''} + {\cal T}_{a''a'''})
\langle  A_{k}^{b}(y) \Pi_{i'}^{a'}(x') A_{i''}^{a''}(x'') \rangle_c \langle
A_{i}^{c}(x) A_{i'''}^{a'''}(x''') \rangle_c]_{(x=y)}
\nn\nn
&&- g (1+ {\cal P}_{ii'})
f^{abc} \partial_{y_i} (1+ {\cal T}_{cb}) [
\langle  A_{k}^{b}(x) \Pi_{i'}^{a'}(x') A_{i''}^{a''}(x'')
A_{i'''}^{a'''}(x''') \rangle_c \langle  A_{k}^{c}(y) \rangle_c
\nn&&+(1+ {\cal T}_{a'a'''} + {\cal T}_{a''a'''})
\langle  A_{k}^{b}(x) \Pi_{i'}^{a'}(x') A_{i''}^{a''}(x'') \rangle_c \langle
A_{k}^{c}(y) A_{i'''}^{a'''}(x''') \rangle_c]_{(x=y)} ,
\label{gluon4ppaa}
\eea
%
\bea
\lefteqn{
\frac{d}{dt} \langle  \Pi_{i}^{a}(x) \Pi_{i'}^{a'}(x') \Pi_{i''}^{a''} (x'')
A_{i'''}^{a'''} (x''') \rangle_c =
 \langle  \Pi_{i}^{a}(x) \Pi_{i'}^{a'}(x') \Pi_{i''}^{a''}(x'')
\Pi_{i'''}^{a'''}(x''') \rangle_c}
\nn\nn
&&+ (1+ {\cal P}_{ii'}+ {\cal P}_{ii''})
[\partial^2_{x}  \langle  A_{i}^{a}(x) \Pi_{i'}^{a'}(x') \Pi_{i''}^{a''}(x'')
A_{i'''}^{a'''}(x''') \rangle_c
\nn &&- \partial_{x_i}  \partial_{x_k}  \langle  A_{k}^{a}(x) \Pi_{i'}^{a'}(x')
\Pi_{i''}^{a''}(x'') A_{i'''}^{a'''}(x''') \rangle_c]
\nn\nn
&&+ 2 g (1+ {\cal P}_{ii'}+ {\cal P}_{ii''})
f^{abc} \partial_{y_k} (1+ {\cal T}_{cb})[
\langle  A_{k}^{b}(x) \Pi_{i'}^{a'}(x') \Pi_{i''}^{a''}(x'')
A_{i'''}^{a'''}(x''') \rangle_c \langle  A_{i}^{c}(y) \rangle_c
\nn&&+(1+ {\cal T}_{a'a'''} + {\cal T}_{a''a'''})
\langle  A_{k}^{b}(x) \Pi_{i'}^{a'}(x') \Pi_{i''}^{a''}(x'') \rangle_c \langle
A_{i}^{c}(y) A_{i'''}^{a'''}(x''') \rangle_c]_{(x=y)}
\nn\nn
&&+ g^2 (1+ {\cal P}_{ii'}+ {\cal P}_{ii''})
f^{abc} f^{cde} (1+ {\cal T}_{eb} + {\cal T}_{ed})
\nn &&\times [
( \langle  A_{k}^{b}(x) \rangle_c \langle  A_{k}^{d}(x) \rangle_c + \langle
A_{k}^{b}(x) A_{k}^{d}(x) \rangle_c )
\langle  A_{i}^{e}(x) \Pi_{i'}^{a'}(x') \Pi_{i''}^{a''}(x'')
A_{i'''}^{a'''}(x''') \rangle_c
\nn &&+(1+{\cal T}_{db}) \lbrace
\langle  A_{k}^{b}(x) \Pi_{i'}^{a'}(x') \rangle_c \langle  A_{k}^{d}(x)
\Pi_{i''}^{a''}(x'') \rangle_c \langle  A_{i}^{e}(x) A_{i'''}^{a'''}(x''')
\rangle_c
\nn &&+(1+{\cal T}_{a'a'''} + {\cal T}_{a''a'''})
\langle  A_{k}^{b}(x) \Pi_{i'}^{a'}(x') \Pi_{i''}^{a''}(x'') \rangle_c
\langle  A_{k}^{d}(x) A_{i'''}^{a'''}(x''') \rangle_c
\langle  A_{i}^{e}(x) \rangle_c
\rbrace
\nn &&+(1+{\cal T}_{a'a''} + {\cal T}_{a'a'''})
\lbrace
\langle  A_{k}^{b}(x)  A_{k}^{d}(x) \Pi_{i'}^{a'}(x') \rangle_c
\langle  A_{i}^{e}(x) \Pi_{i''}^{a''}(x'') A_{i'''}^{a'''}(x''') \rangle_c
\nn &&+
\langle  A_{k}^{b}(x)  A_{k}^{d}(x) \Pi_{i''}^{a''}(x'') A_{i'''}^{a'''}(x''')
\rangle_c
\langle  A_{i}^{e}(x) \Pi_{i'}^{a'}(x') \rangle_c
\rbrace ]
\nn\nn
&&+ g (1+ {\cal P}_{ii'}+ {\cal P}_{ii''})
f^{abc} \partial_{y_k} (1+ {\cal T}_{cb})[
\langle  A_{k}^{b}(y) \Pi_{i'}^{a'}(x') \Pi_{i''}^{a''}(x'')
A_{i'''}^{a'''}(x''') \rangle_c \langle  A_{i}^{c}(x) \rangle_c
\nn&&+(1+ {\cal T}_{a'a'''} + {\cal T}_{a''a'''})
\langle  A_{k}^{b}(y) \Pi_{i'}^{a'}(x') \Pi_{i''}^{a''}(x'') \rangle_c \langle
A_{i}^{c}(x) A_{i'''}^{a'''}(x''') \rangle_c]_{(x=y)}
\nn\nn
&&- g (1+ {\cal P}_{ii'}+ {\cal P}_{ii''})
f^{abc} \partial_{y_i} (1+ {\cal T}_{cb})[
\langle  A_{k}^{b}(x) \Pi_{i'}^{a'}(x') \Pi_{i''}^{a''}(x'')
A_{i'''}^{a'''}(x''') \rangle_c \langle  A_{k}^{c}(y) \rangle_c
\nn&&+(1+ {\cal T}_{a'a'''} + {\cal T}_{a''a'''})
\langle  A_{k}^{b}(x) \Pi_{i'}^{a'}(x') \Pi_{i''}^{a''}(x'') \rangle_c \langle
A_{k}^{c}(y) A_{i'''}^{a'''}(x''') \rangle_c]_{(x=y)} ,
\label{gluon4pppa}
\eea
%
\bea
\lefteqn{
\frac{d}{dt}
\langle  \Pi_{i}^{a}(x) \Pi_{i'}^{a'}(x') \Pi_{i''}^{a''} (x'')
\Pi_{i'''}^{a'''} (x''') \rangle_c =}
\nn
&&
(1+ {\cal P}_{ii'}+ {\cal P}_{ii''}+ {\cal P}_{ii'''})
[ \partial^2_{x} \langle  A_{i}^{a}(x) \Pi_{i'}^{a'}(x') \Pi_{i''}^{a''}(x'')
\Pi_{i'''}^{a'''}(x''') \rangle_c
\nn &&- \partial_{x_i} \partial_{x_k}  \langle  A_{k}^{a}(x) \Pi_{i'}^{a'}(x')
\Pi_{i''}^{a''}(x'') \Pi_{i'''}^{a'''}(x''') \rangle_c ]
\nn\nn
&&+ 2 g (1+ {\cal P}_{ii'}+ {\cal P}_{ii''}+ {\cal P}_{ii'''})
f^{abc} \partial_{y_k} (1+ {\cal T}_{cb})  [
\langle  A_{k}^{b}(x) \Pi_{i'}^{a'}(x') \Pi_{i''}^{a''}(x'')
\Pi_{i'''}^{a'''}(x''') \rangle_c \langle  A_{i}^{c}(y) \rangle_c
\nn&&+(1+ {\cal T}_{a'a'''} + {\cal T}_{a''a'''})
\langle  A_{k}^{b}(x) \Pi_{i'}^{a'}(x') \Pi_{i''}^{a''}(x'') \rangle_c \langle
A_{i}^{c}(y) \Pi_{i'''}^{a'''}(x''') \rangle_c]_{(x=y)}
\nn\nn
&&+ g^2 (1+ {\cal P}_{ii'}+ {\cal P}_{ii''}+ {\cal P}_{ii'''})
f^{abc} f^{cde} (1+ {\cal T}_{eb} + {\cal T}_{ed})
\nn &&\times [
( \langle  A_{k}^{b}(x) \rangle_c \langle  A_{k}^{d}(x) \rangle_c + \langle
A_{k}^{b}(x) A_{k}^{d}(x) \rangle_c )
\langle  A_{i}^{e}(x) \Pi_{i'}^{a'}(x') \Pi_{i''}^{a''}(x'')
\Pi_{i'''}^{a'''}(x''') \rangle_c
\nn &&+(1+{\cal T}_{db}) \lbrace
\langle  A_{k}^{b}(x) \Pi_{i'}^{a'}(x') \rangle_c \langle  A_{k}^{d}(x)
\Pi_{i''}^{a''}(x'') \rangle_c \langle  A_{i}^{e}(x) \Pi_{i'''}^{a'''}(x''')
\rangle_c
\nn &&+(1+{\cal T}_{a'a'''} + {\cal T}_{a''a'''})
\langle  A_{k}^{b}(x) \Pi_{i'}^{a'}(x') \Pi_{i''}^{a''}(x'') \rangle_c
\langle  A_{k}^{d}(x) \Pi_{i'''}^{a'''}(x''') \rangle_c
\langle  A_{i}^{e}(x) \rangle_c
\rbrace
\nn &&+(1+{\cal T}_{a'a''} + {\cal T}_{a'a'''})
\lbrace
\langle  A_{k}^{b}(x)  A_{k}^{d}(x) \Pi_{i'}^{a'}(x') \rangle_c
\langle  A_{i}^{e}(x) \Pi_{i''}^{a''}(x'') \Pi_{i'''}^{a'''}(x''') \rangle_c
\nn &&+
\langle  A_{k}^{b}(x)  A_{k}^{d}(x) \Pi_{i''}^{a''}(x'')
\Pi_{i'''}^{a'''}(x''') \rangle_c
\langle  A_{i}^{e}(x) \Pi_{i'}^{a'}(x') \rangle_c
\rbrace ]
\nn\nn
&&+ g (1+ {\cal P}_{ii'}+ {\cal P}_{ii''}+ {\cal P}_{ii'''})
f^{abc} \partial_{y_k} (1+ {\cal T}_{cb})[
\langle  A_{k}^{b}(y) \Pi_{i'}^{a'}(x') \Pi_{i''}^{a''}(x'')
\Pi_{i'''}^{a'''}(x''') \rangle_c \langle  A_{i}^{c}(x) \rangle_c
\nn&&+(1+ {\cal T}_{a'a'''} + {\cal T}_{a''a'''})
\langle  A_{k}^{b}(y) \Pi_{i'}^{a'}(x') \Pi_{i''}^{a''}(x'') \rangle_c \langle
A_{i}^{c}(x) \Pi_{i'''}^{a'''}(x''') \rangle_c]_{(x=y)}
\nn\nn
&&- g (1+ {\cal P}_{ii'}+ {\cal P}_{ii''}+ {\cal P}_{ii'''})
f^{abc} \partial_{y_i} (1+ {\cal T}_{cb})[
\langle  A_{k}^{b}(x) \Pi_{i'}^{a'}(x') \Pi_{i''}^{a''}(x'')
\Pi_{i'''}^{a'''}(x''') \rangle_c \langle  A_{k}^{c}(y) \rangle_c
\nn&&+( 1+ {\cal T}_{a'a'''} + {\cal T}_{a''a'''})
\langle  A_{k}^{b}(x) \Pi_{i'}^{a'}(x') \Pi_{i''}^{a''}(x'') \rangle_c \langle
A_{k}^{c}(y) \Pi_{i'''}^{a'''}(x''') \rangle_c]_{(x=y)} .
\label{gluon4pppp}
\eea
\subsection{Mixed sector}

The remaining equations of motion - not specified in Section 3 - read as
follows:
%
%
\bea
\lefteqn{
i \frac{d}{dt} \langle \ubr_{\alpha}^{a} (x) u_{\alpha'}^{a'} (x')
A_{i}^{b}(y)  A_{i'}^{b'}(y') \rangle_c = }
\nn &&
  i \langle  \ubr _{\alpha}^{a}(x)  u  _{\alpha'}^{a'}(x')  \Pi _{i}^{b}(y)  A
_{i'}^{b'}(y')  \rangle_c
+ i \langle  \ubr _{\alpha}^{a}(x)  u  _{\alpha'}^{a'}(x')  A _{i}^{b}(y)  \Pi
_{i'}^{b'}(y')  \rangle_c
\nn\nn &&
+  \alpha^{k}_{\beta\alpha}  [
\partial_{x_k}  \langle  \ubr _{\beta}^{a}(x)  u  _{\alpha'}^{a'}(x')  A
_{i}^{b}(y)  A _{i'}^{b'}(y')  \rangle_c
\nn &&
- g  t^{b''ac}  \lbrace
  \langle  \ubr _{\beta}^{c}(x)  u  _{\alpha'}^{a'}(x')  \rangle_c  \langle  A
_{k}^{b''}(x)  A _{i}^{b}(y)  A _{i'}^{b'}(y')  \rangle_c
+ \langle  \ubr _{\beta}^{c}(x)  u  _{\alpha'}^{a'}(x')  A _{i}^{b}(y)
\rangle_c  \langle  A _{k}^{b''}(x)  A _{i'}^{b'}(y')  \rangle_c
\nn &&
+ \langle  \ubr _{\beta}^{c}(x)  u  _{\alpha'}^{a'}(x')  A _{i'}^{b'}(y')
\rangle_c  \langle  A _{k}^{b''}(x)  A _{i}^{b}(y)  \rangle_c
+ \langle  \ubr _{\beta}^{c}(x)  u  _{\alpha'}^{a'}(x')  A _{i}^{b}(y)  A
_{i'}^{b'}(y')  \rangle_c  \langle  A _{k}^{b''}(x)  \rangle_c \rbrace ]
\nn\nn &&
+  \alpha^{k}_{\alpha'\beta}  [
\partial_{x'_k}  \langle  \ubr _{\alpha}^{a}(x)  u  _{\beta}^{a'}(x')  A
_{i}^{b}(y)  A _{i'}^{b'}(y')  \rangle_c
\nn &&
- g  t^{b''a'c}  \lbrace
  \langle  \ubr _{\alpha}^{a}(x)  u  _{\beta}^{c}(x')  \rangle_c  \langle  A
_{k}^{b''}(x')  A _{i}^{b}(y)  A _{i'}^{b'}(y')  \rangle_c
+ \langle  \ubr _{\alpha}^{a}(x)  u  _{\beta}^{c}(x')  A _{i}^{b}(y)  \rangle_c
 \langle  A _{k}^{b''}(x')  A _{i'}^{b'}(y')  \rangle_c
\nn &&
+ \langle  \ubr _{\alpha}^{a}(x)  u  _{\beta}^{c}(x')  A _{i'}^{b'}(y')
\rangle_c  \langle  A _{k}^{b''}(x')  A _{i}^{b}(y)  \rangle_c
\nn &&
+ \langle  \ubr _{\alpha}^{a}(x)  u  _{\beta}^{c}(x')  A _{i}^{b}(y)  A
_{i'}^{b'}(y')  \rangle_c  \langle  A _{k}^{b''}(x')  \rangle_c \rbrace] ,
\label{mixed4aa}
\eea
%
%
\bea
\lefteqn{
i \frac{d}{dt} \langle \ubr_{\alpha}^{a} (x) u_{\alpha'}^{a'} (x')
\Pi_{i}^{b}(y)  A_{i'}^{b'}(y') \rangle_c =
i \langle  \ubr _{\alpha}^{a}(x)  u  _{\alpha'}^{a'}(x')  \Pi _{i}^{b}(y)  \Pi
_{i'}^{b'}(y')  \rangle_c}
\nn &&
+  \alpha^{k}_{\beta\alpha}  [
\partial_{x_k}  \langle  \ubr _{\beta}^{a}(x)  u  _{\alpha'}^{a'}(x')  \Pi
_{i}^{b}(y)  A _{i'}^{b'}(y')  \rangle_c
\nn &&
- g  t^{b''ac} \lbrace
  \langle  \ubr _{\beta}^{c}(x)  u  _{\alpha'}^{a'}(x')  \rangle_c  \langle  A
_{k}^{b''}(x)  \Pi _{i}^{b}(y)  A _{i'}^{b'}(y')  \rangle_c
+ \langle  \ubr _{\beta}^{c}(x)  u  _{\alpha'}^{a'}(x')  \Pi _{i}^{b}(y)
\rangle_c  \langle  A _{k}^{b''}(x)  A _{i'}^{b'}(y')  \rangle_c
\nn &&
+ \langle  \ubr _{\beta}^{c}(x)  u  _{\alpha'}^{a'}(x')  A _{i'}^{b'}(y')
\rangle_c  \langle  A _{k}^{b''}(x)  \Pi _{i}^{b}(y)  \rangle_c
+ \langle  \ubr _{\beta}^{c}(x)  u  _{\alpha'}^{a'}(x')  \Pi _{i}^{b}(y)  A
_{i'}^{b'}(y')  \rangle_c  \langle  A _{k}^{b''}(x)  \rangle_c \rbrace ]
\nn\nn &&
+  \alpha^{k}_{\alpha'\beta}  [
\partial_{x'_k}  \langle  \ubr _{\alpha}^{a}(x)  u  _{\beta}^{a'}(x')  \Pi
_{i}^{b}(y)  A _{i'}^{b'}(y')  \rangle_c
\nn &&
- g  t^{b''a'c} \lbrace
   \langle  \ubr _{\alpha}^{a}(x)  u  _{\beta}^{c}(x')  \rangle_c  \langle  A
_{k}^{b''}(x')  \Pi _{i}^{b}(y)  A _{i'}^{b'}(y')  \rangle_c
+  \langle  \ubr _{\alpha}^{a}(x)  u  _{\beta}^{c}(x')  \Pi _{i}^{b}(y)
\rangle_c  \langle  A _{k}^{b''}(x')  A _{i'}^{b'}(y')  \rangle_c
\nn &&
+  \langle  \ubr _{\alpha}^{a}(x)  u  _{\beta}^{c}(x')  A _{i'}^{b'}(y')
\rangle_c  \langle  A _{k}^{b''}(x')  \Pi _{i}^{b}(y)  \rangle_c
+  \langle  \ubr _{\alpha}^{a}(x)  u  _{\beta}^{c}(x')  \Pi _{i}^{b}(y)  A
_{i'}^{b'}(y')  \rangle_c  \langle  A _{k}^{b''}(x')  \rangle_c \rbrace]
\nn\nn &&
+ i \partial^2_y  \langle  \ubr _{\alpha}^{a}(x)  u  _{\alpha'}^{a'}(x')  A
_{i}^{b}(y)  A _{i'}^{b'}(y')  \rangle_c
- i \partial_{y_i}  \partial_{y_k}  \langle  \ubr _{\alpha}^{a}(x)  u
_{\alpha'}^{a'}(x')  A _{k}^{b}(y)  A _{i'}^{b'}(y')  \rangle_c
\nn\nn &&
+ i  2 g  f^{bb''c}  \partial_{z_k} [
  \langle  \ubr _{\alpha}^{a}(x)  u  _{\alpha'}^{a'}(x')  A _{k}^{b''}(y)
\rangle_c  \langle  A _{i}^{c}(z)  A _{i'}^{b'}(y')  \rangle_c
\nn &&
+ \langle  \ubr _{\alpha}^{a}(x)  u  _{\alpha'}^{a'}(x')  A _{i}^{c}(z)
\rangle_c  \langle  A _{k}^{b''}(y)  A _{i'}^{b'}(y')  \rangle_c
+ \langle  \ubr _{\alpha}^{a}(x)  u  _{\alpha'}^{a'}(x')  A _{i}^{c}(z)  A
_{i'}^{b'}(y')  \rangle_c  \langle  A _{k}^{b''}(y)  \rangle_c
\nn &&
+ \langle  \ubr _{\alpha}^{a}(x)  u  _{\alpha'}^{a'}(x')  A _{k}^{b''}(y)  A
_{i'}^{b'}(y')  \rangle_c  \langle  A _{i}^{c}(z)  \rangle_c  ]_{(y=z)}
\nn\nn &&
+ i g^2  f^{bb''c}  f^{cde} ( 1+ {\cal T}_{b''d}+{\cal T}_{b''e})[
  \langle  \ubr _{\alpha}^{a}(x)  u  _{\alpha'}^{a'}(x')  A _{k}^{b''}(y)
\rangle_c  \langle  A _{k}^{d}(y)  A _{i}^{e}(y)  A _{i'}^{b'}(y')  \rangle_c
\nn &&
+ ( 1+ {\cal T}_{de})
\langle  \ubr _{\alpha}^{a}(x)  u  _{\alpha'}^{a'}(x')  A _{k}^{b''}(y)
\rangle_c  \langle  A _{k}^{d}(y)  \rangle_c  \langle  A _{i}^{e}(y)  A
_{i'}^{b'}(y')  \rangle_c
\nn &&
+ (1+ {\cal T}_{b''d}{\cal T}_{b'e})
\langle  \ubr _{\alpha}^{a}(x)  u  _{\alpha'}^{a'}(x')  A _{k}^{b''}(y)  A
_{i'}^{b'}(y')  \rangle_c  \langle  A _{k}^{d}(y)  A _{i}^{e}(y)  \rangle_c
\nn &&
+ \langle  \ubr _{\alpha}^{a}(x)  u  _{\alpha'}^{a'}(x')  A _{k}^{b''}(y)  A
_{i'}^{b'}(y')  \rangle_c  \langle  A _{k}^{d}(y)  \rangle_c  \langle  A
_{i}^{e}(y)  \rangle_c ]
\nn\nn &&
+ i g  f^{bb''c}  \partial_{z_k}  [
  \langle  \ubr _{\alpha}^{a}(x)  u  _{\alpha'}^{a'}(x')  A _{k}^{b''}(z)
\rangle_c  \langle  A _{i}^{c}(y)  A _{i'}^{b'}(y')  \rangle_c
\nn &&
+ \langle  \ubr _{\alpha}^{a}(x)  u  _{\alpha'}^{a'}(x')  A _{i}^{c}(y)
\rangle_c  \langle  A _{k}^{b''}(z)  A _{i'}^{b'}(y')  \rangle_c
+ \langle  \ubr _{\alpha}^{a}(x)  u  _{\alpha'}^{a'}(x')  A _{i}^{c}(y)  A
_{i'}^{b'}(y')  \rangle_c  \langle  A _{k}^{b''}(z)  \rangle_c
\nn &&
+ \langle  \ubr _{\alpha}^{a}(x)  u  _{\alpha'}^{a'}(x')  A _{k}^{b''}(z)  A
_{i'}^{b'}(y')  \rangle_c  \langle  A _{i}^{c}(y)  \rangle_c  ]_{(y=z)}
\nn\nn &&
- ig  f^{bb''c}  \partial_{z_i}  [
   \langle  \ubr _{\alpha}^{a}(x)  u  _{\alpha'}^{a'}(x')  A _{k}^{b''}(y)
\rangle_c  \langle  A _{k}^{c}(z)  A _{i'}^{b'}(y')  \rangle_c
\nn &&
+  \langle  \ubr _{\alpha}^{a}(x)  u  _{\alpha'}^{a'}(x')  A _{k}^{c}(z)
\rangle_c  \langle  A _{k}^{b''}(y)  A _{i'}^{b'}(y')  \rangle_c
+  \langle  \ubr _{\alpha}^{a}(x)  u  _{\alpha'}^{a'}(x')  A _{k}^{c}(z)  A
_{i'}^{b'}(y')  \rangle_c  \langle  A _{k}^{b''}(y)  \rangle_c
\nn &&
+  \langle  \ubr _{\alpha}^{a}(x)  u  _{\alpha'}^{a'}(x')  A _{k}^{b''}(y)  A
_{i'}^{b'}(y')  \rangle_c  \langle  A _{k}^{c}(z)  \rangle_c  ]_{(y=z)}
\nn\nn  &&
+ i g  t^{ba'c}  \gamma ^{i}_{\alpha'\beta'}  \delta (x'-y)  \langle  \ubr
_{\alpha}^{a}(x)  u  _{\beta'}^{c}(y)  A _{i'}^{b'}(y')  \rangle_c
\nn &&
- ig  t^{bb''c}  \gamma ^{i}_{\beta\beta'}  [
   \langle  \ubr _{\alpha}^{a}(x)  u  _{\beta'}^{c}(y)  A _{i'}^{b'}(y')
\rangle_c  \langle  \ubr _{\beta}^{b''}(y)  u  _{\alpha'}^{a'}(x')  \rangle_c
\nn &&
+  \langle  \ubr _{\alpha}^{a}(x)  u  _{\beta'}^{c}(y)  \rangle_c  \langle
\ubr _{\beta}^{b''}(y)  u  _{\alpha'}^{a'}(x')  A _{i'}^{b'}(y')  \rangle_c ],
\label{mixed4pa}
\eea
%
%
\bea
\lefteqn{
i \frac{d}{dt} \langle \ubr_{\alpha}^{a} (x) u_{\alpha'}^{a'} (x')
\Pi_{i}^{b}(y)  \Pi_{i'}^{b'}(y') \rangle_c =}
\nn &&
\alpha^{k}_{\beta\alpha}  [
\partial_{x_k}  \langle  \ubr _{\beta}^{a}(x)  u  _{\alpha'}^{a'}(x')  \Pi
_{i}^{b}(y)  \Pi _{i'}^{b'}(y')  \rangle_c
\nn &&
- g  t^{b''ac} \lbrace
  \langle  \ubr _{\beta}^{c}(x)  u  _{\alpha'}^{a'}(x')  \rangle_c  \langle  A
_{k}^{b''}(x)  \Pi _{i}^{b}(y)  \Pi _{i'}^{b'}(y')  \rangle_c
+ \langle  \ubr _{\beta}^{c}(x)  u  _{\alpha'}^{a'}(x')  \Pi _{i}^{b}(y)
\rangle_c  \langle  A _{k}^{b''}(x)  \Pi _{i'}^{b'}(y')  \rangle_c
\nn &&
+ \langle  \ubr _{\beta}^{c}(x)  u  _{\alpha'}^{a'}(x')  \Pi _{i'}^{b'}(y')
\rangle_c  \langle  A _{k}^{b''}(x)  \Pi _{i}^{b}(y)  \rangle_c
+ \langle  \ubr _{\beta}^{c}(x)  u  _{\alpha'}^{a'}(x')  \Pi _{i}^{b}(y)  \Pi
_{i'}^{b'}(y')  \rangle_c  \langle  A _{k}^{b''}(x)  \rangle_c\rbrace ]
\nn\nn &&
+  \alpha^{k}_{\alpha'\beta}  [
\partial_{x'_k}  \langle  \ubr _{\alpha}^{a}(x)  u  _{\beta}^{a'}(x')  \Pi
_{i}^{b}(y)  \Pi _{i'}^{b'}(y')  \rangle_c
\nn &&
- g  t^{b''a'c}  \lbrace
  \langle  \ubr _{\alpha}^{a}(x)  u  _{\beta}^{c}(x')  \rangle_c  \langle  A
_{k}^{b''}(x')  \Pi _{i}^{b}(y)  \Pi _{i'}^{b'}(y')  \rangle_c
+ \langle  \ubr _{\alpha}^{a}(x)  u  _{\beta}^{c}(x')  \Pi _{i}^{b}(y)
\rangle_c  \langle  A _{k}^{b''}(x')  \Pi _{i'}^{b'}(y')  \rangle_c
\nn &&
+ \langle  \ubr _{\alpha}^{a}(x)  u  _{\beta}^{c}(x')  \Pi _{i'}^{b'}(y')
\rangle_c  \langle  A _{k}^{b''}(x')  \Pi _{i}^{b}(y)  \rangle_c
+ \langle  \ubr _{\alpha}^{a}(x)  u _{\beta}^{c}(x')  \Pi _{i}^{b}(y)  \Pi
_{i'}^{b'}(y')  \rangle_c  \langle  A _{k}^{b''}(x')  \rangle_c \rbrace ]
\nn\nn &&
+ i ( 1 + {\cal P}_{ii'}) [
\partial^2_y  \langle  \ubr _{\alpha}^{a}(x)  u  _{\alpha'}^{a'}(x')
  A _{i}^{b}(y)  \Pi _{i'}^{b'}(y')  \rangle_c
-\partial_{y_i}  \partial_{y_k}  \langle  \ubr _{\alpha}^{a}(x)  u
_{\alpha'}^{a'}(x')
  A _{k}^{b}(y)  \Pi _{i'}^{b'}(y')  \rangle_c ]
\nn\nn &&
+ i 2g ( 1 + {\cal P}_{ii'}) f^{bb''c}  \partial_{z_k}  [
   \langle  \ubr _{\alpha}^{a}(x)  u  _{\alpha'}^{a'}(x')  A _{k}^{b''}(y)
\rangle_c  \langle  A _{i}^{c}(z)  \Pi _{i'}^{b'}(y')  \rangle_c
\nn &&
+  \langle  \ubr _{\alpha}^{a}(x)  u  _{\alpha'}^{a'}(x')  A _{i}^{c}(z)
\rangle_c  \langle  A _{k}^{b''}(y)  \Pi _{i'}^{b'}(y')  \rangle_c
+  \langle  \ubr _{\alpha}^{a}(x)  u  _{\alpha'}^{a'}(x')  A _{i}^{c}(z)  \Pi
_{i'}^{b'}(y')  \rangle_c  \langle  A _{k}^{b''}(y)  \rangle_c
\nn &&
+  \langle  \ubr _{\alpha}^{a}(x)  u  _{\alpha'}^{a'}(x')  A _{k}^{b''}(y)  \Pi
_{i'}^{b'}(y')  \rangle_c  \langle  A _{i}^{c}(z)  \rangle_c  ]_{(y=z)}
\nn\nn &&
+ i g^2 ( 1 + {\cal P}_{ii'}) f^{bb''c}  f^{cde}
(1+ {\cal T}_{b''d}+ {\cal T}_{b''e})[
 \langle  \ubr _{\alpha}^{a}(x)  u  _{\alpha'}^{a'}(x')  A _{k}^{b''}(y)
\rangle_c  \langle  A _{k}^{d}(y)  A _{i}^{e}(y)  \Pi _{i'}^{b'}(y')  \rangle_c
\nn &&
+ (1+ {\cal T}_{de})
\langle  \ubr _{\alpha}^{a}(x)  u  _{\alpha'}^{a'}(x')  A _{k}^{b''}(y)
\rangle_c  \langle  A _{k}^{d}(y)  \rangle_c  \langle  A _{i}^{e}(y)  \Pi
_{i'}^{b'}(y')  \rangle_c
\nn &&
+ ( 1+ {\cal T}_{b''d}{\cal T}_{b'e})
\langle  \ubr _{\alpha}^{a}(x)  u  _{\alpha'}^{a'}(x')  A _{k}^{b''}(y)  \Pi
_{i'}^{b'}(y')  \rangle_c  \langle  A _{k}^{d}(y)  A _{i}^{e}(y)  \rangle_c
\nn &&
+  \langle  \ubr _{\alpha}^{a}(x)  u  _{\alpha'}^{a'}(x')  A _{k}^{b''}(y)  \Pi
_{i'}^{b'}(y')  \rangle_c  \langle  A _{k}^{d}(y)  \rangle_c  \langle  A
_{i}^{e}(y)  \rangle_c ]
\nn\nn &&
+ i g ( 1 + {\cal P}_{ii'}) f^{bb''c}  \partial_{z_k} [
  \langle  \ubr _{\alpha}^{a}(x)  u  _{\alpha'}^{a'}(x')  A _{k}^{b''}(z)
\rangle_c  \langle  A _{i}^{c}(y)  \Pi _{i'}^{b'}(y')  \rangle_c
\nn &&
+ \langle  \ubr _{\alpha}^{a}(x)  u  _{\alpha'}^{a'}(x')  A _{i}^{c}(y)
\rangle_c  \langle  A _{k}^{b''}(z)  \Pi _{i'}^{b'}(y')  \rangle_c
+ \langle  \ubr _{\alpha}^{a}(x)  u  _{\alpha'}^{a'}(x')  A _{i}^{c}(y)  \Pi
_{i'}^{b'}(y')  \rangle_c  \langle  A _{k}^{b''}(z)  \rangle_c
\nn &&
+ \langle  \ubr _{\alpha}^{a}(x)  u  _{\alpha'}^{a'}(x')  A _{k}^{b''}(z)  \Pi
_{i'}^{b'}(y')  \rangle_c  \langle  A _{i}^{c}(y)  \rangle_c  ]_{(y=z)}
\nn\nn &&
- i g ( 1 + {\cal P}_{ii'}) f^{bb''c}  \partial_{z_i}  [
  \langle  \ubr _{\alpha}^{a}(x)  u  _{\alpha'}^{a'}(x')  A _{k}^{b''}(y)
\rangle_c  \langle  A _{k}^{c}(z)  \Pi _{i'}^{b'}(y')  \rangle_c
\nn &&
+ \langle  \ubr _{\alpha}^{a}(x)  u  _{\alpha'}^{a'}(x')  A _{k}^{c}(z)
\rangle_c  \langle  A _{k}^{b''}(y)  \Pi _{i'}^{b'}(y')  \rangle_c
+ \langle  \ubr _{\alpha}^{a}(x)  u  _{\alpha'}^{a'}(x')  A _{k}^{c}(z)  \Pi
_{i'}^{b'}(y')  \rangle_c  \langle  A _{k}^{b''}(y)  \rangle_c
\nn &&
+ \langle  \ubr _{\alpha}^{a}(x)  u  _{\alpha'}^{a'}(x')  A _{k}^{b''}(y)  \Pi
_{i'}^{b'}(y')  \rangle_c  \langle  A _{k}^{c}(z)  \rangle_c  ]_{y=z)}
\nn\nn &&
+ i g ( 1 + {\cal P}_{ii'}) t^{ba'c}  \gamma ^{i}_{\alpha'\beta'}
\delta (x'-y)  \langle  \ubr _{\alpha}^{a}(x)  u  _{\beta'}^{c}(y)
\Pi _{i'}^{b'}(y')  \rangle_c
\nn &&
- i g ( 1 + {\cal P}_{ii'}) t^{bb''c}  \gamma ^{i}_{\beta\beta'}  [
\langle  \ubr _{\alpha}^{a}(x)  u  _{\beta'}^{c}(y)  \Pi _{i'}^{b'}(y')
\rangle_c  \langle  \ubr _{\beta}^{b''}(y)  u  _{\alpha'}^{a'}(x')  \rangle_c
\nn &&
+ \langle  \ubr _{\alpha}^{a}(x)  u _{\beta'}^{c}(y)  \rangle_c  \langle  \ubr
_{\beta}^{b''}(y)  u  _{\alpha'}^{a'}(x')  \Pi_{i'}^{b'}(y')  \rangle_c ] .
\label{mixed4pp}
\eea
This completes the nonlinear CQCD equations, which form a closed set
of first order differential equations in time and can be integrated
by standard numerical techniques for almost arbitrary initial conditions,
provided that these fulfill the weak Gauss law as well as the Ward identities
as discussed in Section 4.

\end{appendix}

\end{document}